\documentclass[a4paper,dvips,12pt]{article}
\usepackage{epsfig}
\usepackage{amsmath,amssymb,exscale}   
\usepackage{array,multicol}

\newcommand{\NPB}[3]{\emph{ Nucl.~Phys.} \textbf{B#1} (#2) #3}   
\newcommand{\PLB}[3]{\emph{ Phys.~Lett.} \textbf{B#1} (#2) #3}   
\newcommand{\PRD}[3]{\emph{ Phys.~Rev.} \textbf{D#1} (#2) #3}

\def\mco{\multicolumn}

\def\grav{\chi}
\def\ov{\overline}
\def\npa{\not \!\partial}

\def\dalemb#1#2{{\vbox{\hrule height .#2pt
        \hbox{\vrule width.#2pt height#1pt \kern#1pt
                \vrule width.#2pt}
        \hrule height.#2pt}}}

\let\a=\alpha \let\b=\beta   
    
    \let\p=\pi

     \let\F=\Phi

 \def\bd{\begin{document}} \def\ed{\end{document}}
\def\ds{\documentstyle} \let\fr=\frac \let\bl=\bigl \let\br=\bigr
\let\Br=\Bigr \let\Bl=\Bigl 
\let\bm=\bibitem
\let\na=\nabla
\let\pa=\partial 
\let\ov=\overline
\def\ie{{\it i.e.\ }} 
\def\tr{{\mbox{\rm tr}}}
\newcommand{\be}{\begin{equation}}
\newcommand{\ee}{\end{equation}}
\newcommand{\beba}{\begin{equation}\begin{array}{lcl}}
\newcommand{\eaee}{\end{array}\end{equation}}
\newcommand{\bea}{\begin{eqnarray}}
\newcommand{\eea}{\end{eqnarray}}
\newcommand{\ba}{\begin{array}}
\newcommand{\ea}{\end{array}}
\newcommand{\td}{\tilde}
\newcommand{\norsl}{\normalsize\sl}
\newcommand{\ns}{\normalsize}
\newcommand{\refs}[1]{(\ref{#1})}
\def\simlt{\mathrel{\lower2.5pt\vbox{\lineskip=0pt\baselineskip=0pt
           \hbox{$<$}\hbox{$\sim$}}}}
\def\simgt{\mathrel{\lower2.5pt\vbox{\lineskip=0pt\baselineskip=0pt
           \hbox{$>$}\hbox{$\sim$}}}}
\def\A{{\cal A}}
\def\a{{\mathcal a}}
\def\V{{\cal V}}
\def\F{{\cal F}}
\def\p{{\mathcal \phi}}
\def\L{{\mathcal L}}
\def\M{{\mathcal M}}
\def\bD{{\ov {\rm D}}}
\def\bO{{\ov {\rm O}}}
\def\bOp{{\ov {\rm O'}}}

\title{   
\vspace*{-0.8cm}   
\begin{flushright}   
\normalsize{CERN-TH/2001-263\\ CPTH-S039-10-01\\        
\texttt{hep-th/0111209}}\\ 
\end{flushright}    
\vspace{1cm}
\Large\textbf{D-brane models with non-linear supersymmetry}
\author{\large
{\bf I.~Antoniadis~$^1$\footnote{On leave of absence from CPHT,
Ecole Polytechnique, UMR du CNRS 7644.}, K.~Benakli~$^1$ and
A. Laugier~$^{1,2}$}\\ \\
\emph{$^1$CERN Theory Division
  CH-1211, Gen{\`e}ve 23, Switzerland }\\
\emph{$^2$Centre de Physique Th{\'e}orique, Ecole Polytechnique,}\\
\emph{91128 Palaiseau, France}}}
\date{}
\begin{document}
\maketitle
\thispagestyle{empty}
\vspace*{2cm}

\begin{abstract}
We study a class of type I string models with supersymmetry broken
on the world-volume of some D-branes and vanishing tree-level
potential.
Despite the non-supersymmetric spectrum, supersymmetry is non-linearly
realized on these D-branes, while it is spontaneously broken in the
bulk
by Scherk-Schwarz boundary conditions. These models can easily
accommodate
3-branes with interesting gauge groups and chiral fermions. We also
study
the effective field theory and in particular we compute the
four-fermion
couplings of the localized Goldstino with the matter fermions on the
brane.
\end{abstract}
\date

\newpage
\section{Introduction}

One of the most challenging problems in string theory is the mechanism
of supersymmetry breaking. A particularly attractive possibility is by
the process of compactification, using Scherk-Schwarz boundary conditions
\cite{ss}. It consists of imposing non-periodic boundary conditions on
the higher-dimensional fields using an exact (discrete) R-symmetry of the
theory. As a result, the Kaluza-Klein spectrum of different components
of the superfields exhibit mass-shifts according to their R-charges and
the supersymmetry breaking scale is set by the compactification scale
\cite{ablt,Kounnas}. This was one of the main motivations for large internal
dimensions in string theory with size of order TeV$^{-1}$ \cite{a}.
Their phenomenological consequences have been extensively studied during 
the last years \cite{Pomarol}.

Type I string theory offers new realizations of the Scherk-Schwarz
mechanism, due to the presence of D-branes. Indeed, for D-branes
transverse to the direction used by the Scherk-Schwarz deformation, the
massless spectrum on their world-volume remains supersymmetric (at
lowest order), while supergravity is spontaneously broken in the bulk
\cite{ads1,ads4}. This is the phenomenon of ``brane supersymmetry", present
also when breaking supersymmetry in M-theory \cite{mth} along the eleventh
dimension \cite{aq}. It is a generalisation of a similar effect present
in orbifold compactifications of the heterotic string, where twisted
fields localized at the orbifold fixed points do not feel the
supersymmetry breaking.

Another attractive possibility in type I string model building is the
mechanism of ``brane supersymmetry breaking" \cite{ads2} generalising   
the stable non-BPS brane construction of refs. \cite{sen,Sug}. Supersymmetry
is now primordially broken on the world-volume of some D-branes while
it remains unbroken (to lowest order) in the closed string bulk.
If the Standard Model is localized on these non-supersymmetric
D-branes, the string scale must be at TeV energies in order to protect
the gauge hierarchy. The observed weakness of gravity can then be
explained by introducing some extra large dimensions in the bulk, of size
that can reach a millimeter \cite{add}. In this case, the spectrum
localized on these D-branes is not supersymmetric, but still
supersymmetry is non-linearly realized \cite{VA}. In particular, there is a
massless Goldstino, localized on their world-volume, transforming
non-linearly under supersymmetry \cite{dm}. On the other hand, radiative
corrections are expected to generate a ``tiny" supersymmetry breaking in
the bulk which should vanish in the large volume limit. One of the main
difficulties of these models is the presence of a non-vanishing
tree-level potential giving rise to tadpoles for the dilaton and other
NS-NS (Neveu-Schwarz) scalars, which make any quantitative study of
these models questionable.

In this work, we construct a class of models exhibiting brane
supersymmetry breaking with vanishing (global) tadpoles. This is
achieved by introducing at the same time Scherk-Schwarz boundary
conditions along some directions orthogonal to the non-supersymmetric
branes, which produce an additional (small) supersymmetry breaking in the
bulk, that vanishes in the decompactification limit. As we show, in
explicit examples, these models can easily accommodate 3-branes with
interesting gauge groups and chiral fermions. Thus, supersymmetry is
broken at the string scale on the branes, although it remains
non-linearly realized, while supergravity is spontaneously broken at the
compactification scale in the bulk. If the string scale is at the TeV
\cite{add,Lykken},
the gravitino mass varies between $10^{-3}$ eV to 10 MeV, for two up to
six extra large transverse dimensions, respectively.

We also study the effective interactions of the Goldstino with the
matter fields living on the brane. In particular, we focus on the
dimension eight four-fermion terms which can be arranged in two
different operators. These were analysed in the past in refs. \cite{Brig,CLL}
using non-linear supersymmetry, that fixes the coupling of one particular
operator but leaves the other undetermined. In this work, we perform a
string computation of both couplings and we determine in particular the
unknown coefficient. In the simplest case, where matter fermions correspond to open strings with both ends on the same set of branes, we find that the Goldstino decay constant is given
by the total tension of the corresponding D-branes,
while the unknown coefficient has two possible values, depending on
whether the matter fermions have the same or different internal helicity with the Goldstino.

The paper is organised as follows. In section 2, we present the effective
field theory describing the Goldstino interactions and we determine the
corresponding couplings by a string computation, using the results of
ref. \cite{Contact}. In section 3, we describe the essential points of our
construction using three ingredients: Scherk-Schwarz deformation in the
bulk, orientifold planes at the boundaries and appropriate introduction
of D-branes on top of the orientifolds. We also discuss the general
properties of the resulting models. In section 4, we present the simplest
model in nine dimensions, using D8-branes, which contains however the
essential properties of our construction. In section 5, we give a
six-dimensional chiral example using D9 and D5 branes. Supersymmetry is
non-linearly realized on the world-volume of D5-branes, while is
spontaneously broken by the Scherk-Schwarz boundary conditions in the
closed string sector and on the D9-branes in the bulk. In section 6, we
present a four-dimensional example with chiral fermions and Pati-Salam 
 type gauge group, having the same characteristics with the
six-dimensional example. Finally, section 7 contains our conclusions.
The results of sections 4,5 and 6 are summarized at the end of section 3,
so that the reader who is not familiar with string theory can skip them
and go directly to the conclusions. We have also included four
appendices containing notations and technical details used in sections 4-6.

\section{Interaction of Goldstino with matter}

The spontaneous breaking of  continuous symmetries lead to the appearance
of Nambu-Goldstone fields. These represent the parameters of (non-linear)
transformations realizing  the broken symmetries in the vacuum.
For the case of supersymmetry, the Nambu-Goldstone field $\grav^\alpha$
has spin-1/2 and  is called  Goldstino \cite{VA}. If the broken
supersymmetry is local then the Goldstino mixes with the gravitino,
giving rise to the longitudinal components of  the massive spin-3/2
particle.

In this section, we describe the leading interaction terms involving two
Goldstinos with two matter fermions.  We will restrict our analysis to
the case of ${\cal N}=1$ supersymmetry in four dimensions, the
generalisation to other dimensions or higher supersymmetries being
straightforward. In superstring models, these amplitudes are easy to
compute and allow to extract the value of the Goldstino decay constant.
On the other hand, the non-observation of effects corresponding to the
production of two Goldstinos  in the interaction of matter fermions has
been used  to derive experimental bounds on the supersymmetry breaking
scale for a class  of models  where the gravitino mass is much smaller
than all other sparticle masses \cite{Zwirner}. In fact, in this case,  
at low energy,
the interactions of gravitinos with matter are dominated by the
interaction with their  spin $1/2$ components, i.e. the Goldstino.

The breaking of supersymmetry is characterized on the one hand by an
order parameter $v$, fixing the Goldstino decay constant, associated to
the primordial breaking scale, and on the other hand by the
mass-splittings of the supermultiplets, $\Delta m_i^2=\lambda_i v^2$,
with $\lambda_i$ the corresponding superpotential couplings.
In the case of spontaneous breaking by a $D$-term, $v^2=D$,  while for a
breaking through a non-vanishing $F$-term, $v^2=\sqrt{2} F$. In the
low energy limit, when all mass-splittings go to infinity with $v$ fixed,
the non-linear supersymmetry transformations of the
Goldstino fields are given by:
\bea
\delta^{SUSY}\, \grav^\alpha&=&
v^2 \, \xi^\alpha -\frac{i}{v^2}\left( \grav^\beta \sigma^\mu
\bar\xi_{\dot\beta}
    -\xi_{\dot\beta} \sigma^\mu \bar\grav^{\dot\beta} \right)
\partial_\mu \grav^\alpha~~\cr
\delta^{SUSY}\bar{\grav}_{\dot\alpha}&=&
v^2 \, \bar\xi_{\dot\alpha}
-\frac{i}{v^2}\left( \grav^\beta \sigma^\mu \bar\xi_{\dot\beta}
    -\xi_{\dot\beta} \sigma^\mu \bar\grav^{\dot\beta} \right)\partial_\mu
\bar\grav_{\dot\alpha}
\eea
where $\xi^\alpha,\bar{\xi}_{\dot\alpha}$
are (two-component) spinorial transformation parameters.

The lowest dimensional operators describing the interaction of two matter
fermions $f$ with two Goldstinos $\grav$ have dimension eight and can be
written as \cite{Brig,CLL}:
\bea
\L = \int d^4x \left[-\frac{1}{2v^4}
\left( \grav \stackrel{\leftrightarrow}{\partial}_\mu
\sigma^\nu \bar\grav\right) \left( f
\stackrel{\leftrightarrow}{\partial}_\nu \sigma^\mu \bar f \right)
+ \frac{2 \, C_{f}}{v^4}
\left( f\partial^\mu \grav\right)\left(\bar
f \partial_\mu \bar\grav\right) \right].
\label{leff}
\eea
The first term corresponds to the coupling of the energy-momentum tensor
of matter fermions with the Goldstino and is model independent. Its
normalization is fixed by the choice of canonical kinetic terms for
the matter fermions. In contrast, the real coefficient $C_{f}$ is  model
dependent\footnote{The parameter $C_f$  is related to the parametrisation of other authors by
$C_f= -\frac{\alpha}{4} \cite{Brig} = \frac{C_{ff}}{2} \cite{CLL}$.} and
describes a possible coupling of matter fermions to a  non-trivial
torsion term of the Goldstino \cite{Brig}. While in the low energy
effective field theory context, the value of $C_f$ is an arbitrary
unknown parameter, it can be computed explicitly in the fundamental
(string) theory.

The values of the two parameters $v$ and $C_{f}$ can be
extracted from the analysis of the amplitude $f \bar f \rightarrow
\grav \bar\grav$. The four-dimensional
momenta corresponding to the particles $f, \bar f, \grav, \bar\grav$
are chosen all directed inward and denoted by $k_i$. We use
the Mandelstam variables (kinematical invariants)
$\{s, t, u\}$ defined as:
\bea
s= -(k_1+k_2)^2 \, ,  \qquad
t= -(k_2+k_3)^2 \, ,  \qquad
u=-(k_1+k_3)^2 .
\eea
The different helicity amplitudes for the process $f\bar f \rightarrow
\grav \bar\grav$ can be easily extracted from the effective Lagrangian
(\ref{leff}):
\bea
\label{agen}
\A (f_L \bar f_R \rightarrow \grav_L \bar\grav_R) &=&
- \frac {2} {v^4} \left( t  \, u +  {C_{f}}  \,  s  \, u \right) \, ,
   \cr
\A(f_L \bar f_R \rightarrow \grav_R \bar\grav_L) &=&
\frac {2} {v^4} \left(  t  \, u + {C_{f}} \,  s \,  t \right) \, ,
\eea
where the subscripts $L$ and $R$ label the left-handed
and right-handed  helicities respectively.

In the following, we consider a scenario where matter fields are
localized on a collection of non supersymmetric D-branes embedded in a
higher-dimensio\-nal bulk. We will work in the limit where the volume of
the bulk (transverse volume) is large compared to the string length
$l_s\equiv M_s^{-1}$. As we discuss below, in this limit, the effects of
supersymmetry breaking in the bulk are negligeable and we can restrict
our analysis to the breaking of global supersymmetry on the brane.

As we describe in sections 4-6, supersymmetry breaking on the world-volume
of D-branes is achieved by placing them together with appropriate
orientifold planes and applying the corresponding orbifold projections.
In the simplest case and in the large transverse volume limit, the gauge
group is orthogonal (symplectic) while fermions transform in the
symmetric (antisymmetric) representation. This representation is
reducible and its singlet component can be identified with the Goldstino
\cite{dm,Bagger,yoneya}. Thus, the spectrum is not supersymmetric on the brane,
although there is exact supergravity in the bulk, but supersymmetry is
non-linearly realized with a massless localized Goldstino. At finite
volume, the gravitino obtains a mass via Scherk-Schwarz boundary
conditions, and the gauginos appear as odd open string winding modes,
transforming in the adjoint representation of the gauge group; in the
large volume limit, they become superheavy and decouple from the
spectrum.

We will consider a system of at most two type of branes:
 Dp-branes and Dq-branes with $p-q =0 \,\,{\rm mod}\,\, 4$. In particular we will
deal with systems of D9 branes or D9--D5 branes which can be mapped 
through $T$-duality to a system of D3 or D7--D3 branes. The six-dimensional
 internal 
space is  compactified on a 
six-dimensional torus. The world-volume of a D$p$-brane
is here extending along the four-dimensional Minkowski space
as well as along a volume $V_p$ in the internal compact space\footnote{Here $V_p$ denotes the volume along the $p-3$ compact directions of the $D_p$-brane.}. 
For the sake of simplicity, we will ignore here 
the effects of the presence of orbifolds and orientifolds which lead 
to model dependent projections of some (or all) goldstinos away from the 
massless spectrum.

In these models, the Goldstino is identified with a massless mode of an
open string with both ends located on parallel D-branes. Such strings
are denoted as ``DD strings'' as the associated  world-sheet fields
satisfy DD (Dirichlet-Dirichlet) boundary conditions along all
transverse directions. If instead the open strings are stretched between
non-parallel branes (or between D$p$ and D$q$ branes with $p \neq q$), they are denoted as ``ND strings'' as the the
associated  world-sheet fields satisfy now ND (Neumann-Dirichlet)
boundary conditions. Their massless modes appear as localized states
living at the corresponding brane intersections.

The scattering amplitudes of open string modes are described, at the
lowest order, by correlation functions of the corresponding vertex
operators on a two-dimensional surface with the topology of a disk.
Each end of the open string carries a charge which is representing the
transformation under the gauge symmetry group $G =U(N_p),\, SO(N_p)$ or $
USp(N_p)$, where $N_p$ is given by the number of parallel D$p$-branes
stacked together. The transformation of the vertex operators under the
gauge group $G$ is encoded in Chan-Paton matrices $\lambda$.
The form of vertex operators is given for instance in ref.
\cite{Contact}, where we choose the normalisation of Chan-Paton matrices
of matter fields as $Tr (\lambda^a\lambda^b) =\delta_{ab}$. Since the 
Goldstinos are gauge
singlets, their corresponding Chan-Paton matrices are
$\lambda^{(a)}_p=\frac{{\bf 1}_{N_p}}{\sqrt{N_p V_p}}$
 with
${\bf 1}_{N_p}$ the identity matrix of rank $N_p$.

On the world-volume of each brane, the space-time fermions transform as 
spinors of $SO(10)$ and they are labeled by their helicity with respect 
to the maximal subgroup 
$SO(2)^5$ as $\frac{1}{2}(\pm \pm \pm \pm \pm)$, the first two (three) 
$SO(2)$'s 
represent the four-dimensional  space-time 
(six-dimensional D5-brane worldvolume) helicities.\footnote{In fact, the space-time fermions transform as 
spinors  of $SO(1,9)$, but it is easier to work in the Euclidean version.}
 Each D$p$-brane breaks  
spontaneousely half 
of the bulk supersymmetries. The corresponding Goldstinos are the gauginos 
of the $U(1)$ gauge boson appearing on the world-volume of the brane.
In the case of $N_p$ parallel branes, the  Goldstinos are the gauginos 
of the overall $U(1)$ with the Chan-Paton matrices given above. In our 
computations we choose the
Goldstino helicity to be $\frac{1}{2}(+++++)$.

We will first consider the case where there is only one type of brane 
and compute the four-fermion interaction involving two Goldstinos and 
two matter fermions. This is obtained from the correlation function of four 
vertex operators representing the emission (absorbtion) of DD open strings
associated with the two Goldstinos and two fermions. For 
the two Goldstinos and conjugate vertices we choose the helicities
$\frac{1}{2}(+++++)$ and $\frac{1}{2}(+----)$ which leaves us with two choices
for the matter fermions helicities:
\begin{itemize}

\item case $(I)_{DD}$ where the fermions 
having six-dimensional internal space  helicity as the Goldstinos, corresponding to 
$\frac{1}{2}(--+++)$ and $\frac{1}{2}(-+---)$.

\item case $(II)_{DD}$ where, with respect of the Goldstinos, the fermions  
have opposite helicity in the internal six-dimensional space, 
corresponding for instance to 
$\frac{1}{2}(-+++-)$ and $\frac{1}{2}(----+)$.

\end{itemize}
Here, the (four-dimensional) space-time helicities are chosen according to the chiralities of the field theory amplitudes (\ref{agen}) that we want to study.

The corresponding scattering amplitudes  are straightforward to obtain. They 
can be computed for instance using the results of ref. \cite{peskin,Contact}. 
The total scattering  
amplitude $\A_{total}(1,2,3,4)$ is
obtained by summing over the six  possible ordered amplitudes: 
$A(1,2,3,4)$ and the five other permutations of the vertex 
operateurs .
It is useful to define $\A(1,2,3,4) = A(1,2,3,4) + A(4,3,2,1)$ as the two amplitudes have the same traces on Chan-Paton matrices, so that:
\bea
\label{totamp}
\A_{total}(\tilde{1},\tilde{2},3,4) = \A (\tilde{1},\tilde{2},3,4) + \A (\tilde{1},3,\tilde{2},4) + \A (\tilde{1},\tilde{2},4,3)
\eea
where the tilde stands for Goldstinos.

\begin{itemize}
\item In the $(I)_{DD}$ configuration
\bea
\label{orderedDD2}
\A(\tilde{1},\tilde{2},3,4)&=&-2g_sl_s^2tr(\tilde{\lambda}^1\tilde{\lambda}^2\lambda^3\lambda^4+\lambda^4\lambda^3\tilde{\lambda}^2\tilde{\lambda}^1)\int_0^1 dx x^{-1-sl_s^2}(1-x)^{-1-tl_s^2}\nonumber \\ &&({\tilde{\bar v}}^{(1)} \gamma_{\mu}\tilde{u}^{(2)} 
{\bar v}^{(4)} \gamma^{\mu} u^{(3)}   (1-x)  -
 {\tilde{\bar v}}^{(1)} \gamma_{\mu} v^{(4)}  
{\bar u}^{(3)} \gamma^{\mu} \tilde{u}^{(2)} x ) 
\eea
which leads after performing the sum over the permutations to:
\bea
\label{IDDtot}
{\cal A}(f_L \overline{f}_R \rightarrow \grav_L \overline{\grav}_R)=-\frac{4}{N_pV_p}g_s\left[(-\frac{2t}{s}-\frac{2t}{t}){\cal F}(s,t)+\frac{2t}{t}{\cal F}(t,u)+\frac{2t}{s}{\cal F}(u,s)\right] \nonumber \\
\simeq -\frac{4\pi^2}{N_pV_pM_s^4}g_su^2 \qquad\qquad\qquad\qquad\qquad\qquad\qquad\qquad\qquad
\eea
where $g_s$ is the type $I$ string coupling and $M_s=l_s^{-1}$ is the string scale. The symbol ``$\simeq$'' means that we have taken the leading order for $|sl_s^2|,|tl_s^2|,|ul_s^2|<<1$ which corresponds to the field theory limit
and used:
\bea
{\cal{F}}(x,y) \simeq 1-\frac{\pi^2}{6}\frac{xy}{M_s^4}\, .
\eea

\item In the $(II)_{DD}$ case we have instead
\bea
\label{orderedDD1}
\A(\tilde{1},\tilde{2},3,4)&=&-2g_sl_s^2tr(\tilde{\lambda}^1\tilde{\lambda}^2\lambda^3\lambda^4+\lambda^4\lambda^3\tilde{\lambda}^2\tilde{\lambda}^1)\int_0^1 dx x^{-1-sl_s^2}(1-x)^{-1-tl_s^2}\nonumber \\ &&(\tilde{{\bar v}}^{(1)} \gamma_{\mu} \tilde{u}^{(2)}
{\bar v}^{(4)} \gamma^{\mu} u^{(3)}   (1-x)  -
 2\tilde{\bar v}^{(1)} v^{(4)}  
{\bar u}^{(3)}\tilde{u}^{(2)} x ) 
\eea
which leads to the total amplitude 
\bea
\label{IIDDtot}
{\cal A}(f_L \overline{f}_R \rightarrow \grav_R \overline{\grav}_L)=-\frac{4}{N_pV_p}g_s\left[(-\frac{2t}{s}-\frac{2t}{t}){\cal F}(s,t)+\frac{2t}{t}{\cal F}(t,u)+\frac{2t}{s}{\cal F}(u,s)\right] \nonumber \\
\simeq \frac{4\pi^2}{N_pV_pM_s^4}g_sut \qquad\qquad\qquad\qquad\qquad\qquad\qquad\qquad\qquad
\eea
\end{itemize}

We observe that there are no poles (only presence of contact terms) and the 
dominant contribution comes from dimension eight operator as expected.
Comparing eqs. (\ref{IDDtot}) and (\ref{IIDDtot}) with eq. (\ref{agen}), we
can  identify $v^4$ and the coefficients $C_f^{DD,I}$ and $C_f^{DD,II}$
corresponding to fermions from DD strings with the same or different six-dimensional internal helicities with the Goldstinos, respectively. We obtain:
\bea
C_f^{DD, I}=1\, , \qquad  C_f^{DD,II} =0 \, , \qquad \frac {v^4}{2} =
N_p V_p\frac{M_s^4}{4 \pi^2 g_s}= N_p\, V_p \, \tau_3\, ,
\eea
where we have identified the D3-brane tension
$\tau_3 =\frac{M_s^4}{4 \pi^2 g_s}$. By performing an appropriate T-duality 
transformation along the directions transverse to the four space-time directions, 
the volume factor $V_p$ disapears and thus D$p$-branes can be viewed as 
$D3$-branes. Thus, in four dimensions, the
Goldstino decay constant is identified with the D3-brane tension times
the RR (Ramond) charge $N_p$, while the parameter $C_f$ takes two possible
values, depending on whether the fermions have the same or oposite internal 
helicity than the Goldstinos.

We consider now a system with two types of branes, for instance D5 or 
 D$\overline{5}$ and D9 
branes. We can decompose the $SO(2)^5$ helicities as six-dimensional and 
internal through $SO(2)^5 = SO(2)^3 \otimes SO(2)^2$, such that 
the helicity in the six-dimensional 
D5 or D$\overline{5}$ branes world-volume is given by the product of the  first three signs. Each type of branes breaks spontaneously 16
supersymmetries out of the 32 present in the bulk. The
corresponding Goldstinos  can be assembled into
two sets of 8-component spinors having the same six-dimensional and 
internal helicities:
\bea
{\rm D}9 \rightarrow 8_{++}+8'_{--}\nonumber \\
{\rm D}5 \rightarrow 8_{++}+8''_{+-}\nonumber \\
{\rm D}\overline{5} \rightarrow 8'''_{-+}+8'_{--}  
\eea
where the first and  second signs correspond to the six-dimensional
and internal helicities, respectively. Note that the D5 and D$\overline{5}$ 
branes together break all supersymmetries as we retrieve all the 32 supercharges among the Goldstinos. On the other hand the D5--D9 or D$\overline{5}$--D9 systems preserve 8 supercharges each. In each of these
two cases,  8 supercharges are broken by both branes and the corresponding 
Goldstinos appear as $U(1)$ gauginos in six dimensions.

We consider the case of Goldstino with helicity $\frac{1}{2}(+++++)$, thus 
among the $8_{++}$ spinors. The Goldstino $\grav$ cannot be only seen as a $55$ or $99$ state associated with open strings with both ends on D5 or D9 branes respectively, but it is 
a linear combination of the two. It is given by:
\bea 
\grav = \sqrt{\frac{N_9V_9}{N_5V_5+N_9V_9}} \grav_9 + 
\sqrt{\frac{N_5V_5}{N_5V_5+N_9V_9}} \grav_5
\eea
Only this combination leads only to dimension eight operators. The 59 strings 
have no charges under the corresponding linear combination for $U(1)$ gauge
 bosons.

In a similar way than the case of a single type of brane, we will choose for 
the Goldstinos and for its conjugate, the helicities
$\frac{1}{2}(+++++)$ and $\frac{1}{2}(+----)$. There are then two cases:
\begin{itemize}

\item case $(I)_{ND}$ where the fermions 
have six-dimensional  space-time  helicity opposite to the one of the 
Goldstinos and  correspond to 
$\frac{1}{2}(-++)$ and $\frac{1}{2}(---)$.

\item case $(II)_{ND}$ where the fermions  have the same
six-dimensional space-time helicities with the Goldstinos, and 
correspond to 
$\frac{1}{2}(--+)$ and $\frac{1}{2}(-+-)$.

\end{itemize}

A straightforward computation leads to:
\begin{itemize}
\item in the $(I)_{ND}$ one
\bea
{\cal A}(f_{L,Ii} \overline{f}_{R,Jj} \rightarrow \grav_R \overline{\grav}_L)&=&
\frac{8}{N_5V_5+N_9V_9}g_s\delta_{IJ}\delta_{ij} \nonumber \\
&\times&\left[(\frac{2t}{s}-\frac{t}{t}){\cal F}(s,t)-\frac{2t}{s}{\cal F}(s,u)+\frac{t}{t}{\cal F}(t,u)\right]\nonumber \\
&\simeq& -\frac{4\pi^2}{(N_5V_5+N_9V_9)M_s^4}g_s\delta_{IJ}\delta_{ij}t^2
\eea

\item for the $(II)_{ND}$ case
\bea
\label{NDnsusy}
{\cal A}(f_{L,Ii} \overline{f}_{R,Jj} \rightarrow \grav_L \overline{\grav}_R)&=&
-\frac{16\pi^2}{3(N_5V_5+N_9V_9)M_s^4}g_s\delta_{IJ}\delta_{ij}\nonumber \\ &\times& [B(-sl_s^2,1-tl_s^2) + B(1-sl_s^2,\frac{1}{2}-tl_s^2) \nonumber \\ &-&B(-sl_s^2,1-ul_s^2)-B(1-ul_s^2,\frac{1}{2}-tl_s^2)] \nonumber \\
&\simeq&-\frac{16\pi^2}{3(N_5V_5+N_9V_9)M_s^4}(1-w)g_s\delta_{IJ}\delta_{ij} \nonumber \\ && \left[ut+\frac{\frac{1}{2}+2w}{1-w}us\right]
\eea
\end{itemize}
where $w=\frac{3\delta}{\pi^2}\simeq0.373$ 
(see below for the definition of $\delta$) and the function $B(a,b)$ is 
defined as :
\bea
B(a,b)=\int_0^1 dx x^{a-1}(1-x)^{b-1}=\frac{\Gamma(a)\Gamma(b)}{\Gamma(a+b)}.
\eea
It is related to the string form factor ${\cal F}(s,t)$ by :
\bea
B(-sl_s^2,-tl_s^2)=\frac{u}{stl_s^2}{\cal F}(s,t) \nonumber \\
B(-sl_s^2,1-tl_s^2)=-\frac{1}{sl_s^2}{\cal F}(s,t)
\eea
and is expanded in Taylor series  around $x,y=0$ as:
\bea
B(1-x,\frac{1}{2}-y)\simeq 2+4y+\delta x\nonumber \\
B(-x,1-y)\simeq-\frac{1}{x}+\frac{\pi^2}{6}y
\eea
where $\delta$ is given  in term of the Euler's constant 
$\gamma\simeq 0.577$ and digamma function value
 $\psi(\frac{3}{2})\simeq0.036$ with $\psi(z)=\frac{\Gamma'(z)}{\Gamma(z)}$:
\bea
\delta=2(\gamma+\psi(\frac{3}{2}))\simeq1.227
\eea
Comparing with the field theory result, we thus get:
\bea
\label{FthvCND}
\frac{v_{ND,I}^4}{2}&=&(N_5V_5+N_9V_9)\tau_3 \,\,\,\,\,\,\,\,\,\,\,\,\,\,\,\,\,\,\,\,\,\,\,\,\,\,\,\,\,\,\,\,\,\,C_f^{ND,I}\,\,\,=\,\,\,1 \nonumber \\ 
\frac{v_{ND,II}^4}{2}&=&\frac{3}{4(1-w)}(N_5V_5+N_9V_9)\tau_3\,\,\,\,\,\,\,\,C_f^{ND,II}\,\,\,=\,\,\,\frac{\frac{1}{2}+2w}{1-w} 
\eea
Note that in the second case, where the matter fermions have the same six-dimensional space-time helicities with the Goldstino, one obtains a different result for both the decay constant and $C_f$. We believe that this is due to the effects of the width of the brane intersection which becomes important in this case.

\section{ Building non-supersymmetric brane-worlds}

Before providing explicit examples in diverse dimensions,
we will present here the main ingredients of the construction and the
main properties of string models with non-linear brane supersymmetry
and vanishing (tree-level) cosmological constant.

\subsection{Construction}

To make the construction simple, we consider one large extra dimension
(used by the Scherk-Schwarz deformation), and  follow three
 steps: (i) coordinate dependent  compactification on
$S^1/{\mathbb{Z}_2}$; (ii) introduction of orientifold planes at the
boundaries of the segment $S^1/{\mathbb{Z}_2}$; (iii) placing D-branes
on top of  the orientifolds.

\vspace{0.5cm}

\noindent {\bf  (i) {\it  Coordinate dependent compactification on
$S^1/{\mathbb{Z}_2}$}}

\vspace{0.3cm}

Although the results on the
Scherk-Schwarz breaking of supersymmetry presented here are known
\cite{ss}, we recall their main features in order to make this section
self-contained.

We start with a supersymmetric string vacuum in $D+1$ dimensions. The
space-time is spanned\footnote{For curved
indices in $D+1$ dimensions we use capital letters
from the middle of  Latin alphabet $M= 0,\cdots,D$, while
letters from the beginning of the alphabet are used for flat
tangent space coordinates. We use corresponding Greek letters
for $D$-dimensional indices.} by the coordinates $X^M$.
The $(D+1)$-dimensional theory has a linearly realized supersymmetry
between the vielbein $e_A^M$  and the gravitino and
between  bosons $\Phi$ and their spin $1/2$
fermionic partners $\Psi$ which can be formally written as:
\bea
\delta_{\eta} \Phi  &\sim&  \bar\eta \Psi  \cr
\delta_{\eta} \Psi  &\sim& \eta {\npa} \Phi \cr
\delta_{\eta} e_a^{M}
&\sim&  \bar\eta \Gamma_a \Psi^{M}  \cr
\delta_{\eta} \Psi^{M}
&\sim& D^{M}\eta  +\cdots
\label{hsusy}
\eea
where $\eta(x^M)$ are local supersymmetry transformation parameters and
$D^M$ is  the appropriate covariant derivative. The dots represent
contributions of other fields ($p$-forms) that are needed to close the
supersymmetry algebra  in $D+1$ dimensions. In eq. (\ref{hsusy})
we neglected bilinear and higher order terms in the fermions.

The space-like $x^D = y $ coordinate is compactified
on a segment $S^1/\mathbb{Z}_2$ of length $\pi R$.
The bosonic fields $\Phi$ and $e_a^M$ satisfy periodic conditions
along $S^1$ while the fermions are anti-periodic:
\bea
y &\rightarrow& y+2\pi R \, : \cr
\{\Phi, e_a^M,\cdots \} (x^\mu, y) &=& \, \,  \,
\{\Phi, e_a^M,\cdots \}  (x^\mu, y+2\pi R)\cr
\{\Psi, \Psi^M \} (x^\mu, y) &=&\! \! \! - \, \{\Psi, \Psi^M \}
(x^\mu, y+2\pi R)
\label{periodic}
\eea
where the dots stand for the additional $p$-form fields.

Under the $\mathbb{Z}_2$ orbifold action, the bosonic fields can be
decomposed into even (labeled by ${e}$) and odd parts (labeled by ${o}$):
\bea
y &\rightarrow& -y \, : \cr
\{\Phi_{e}, e_a^\mu, e_D^D \} (x^\mu, y) &=& \, \,  \,
\{\Phi_{e}, e_a^M, e_D^D \} (x^\mu, -y)\cr
\{\Phi_{o}, e_D^\mu, e_\mu^D \} (x^\mu, y) &=& -
\{\Phi_{o}, e_D^\mu, e_\mu^D \}(x^\mu, -y)
\eea
where the complex scalar fields $\Phi$ have been splitted into even
and odd parts: $\Phi = \Phi_{e} + \Phi_{o}$. The fields
$ \Phi_{e}$ and $\Phi_{o}$ can be decomposed into Fourier modes:
\bea
\Phi_{e}(x^\mu,y) &=& \sum_{n=0}^\infty \Phi^{(n)}_{e}(x^\mu)
\cos{(\frac{n}{R} y)}\cr
\Phi_{o}(x^\mu,y) &=& \sum_{n=0}^\infty \Phi^{(n)}_{o}(x^\mu)
\sin{(\frac{n}{R} y)}
\eea
and similarly for the vielbein and the various $p$-forms. The modes
$n=0$ correspond to massless states which remain
in the $D$-dimensional effective Lagrangian.

The spinors $\Psi$ and $\Psi^M$ are decomposed into two components
in $D$ dimensions:
\bea
\Psi = \left(\begin{array}{l}\psi_{e}\\ \psi_{o}\end{array}
\right);\qquad
\Psi^\mu = \left(\begin{array}{l}\psi^\mu_{e}\\ \psi^\mu_{o}\end{array}
\right);\qquad
\Psi^D = \left(\begin{array}{l}\psi^D_{o}\\ \psi^D_{e}\end{array}
\right)
\eea
which satisfy the following parity transformations:
\bea
y &\rightarrow& -y \, : \cr
\left(\! \!\begin{array}{l}\{\psi_{e}, \psi^\mu_{e}\}, \{\psi^D_{o} \}
\\ \{\psi_{o}, \psi^\mu_{o}\}, \{\psi^D_{e} \}\end{array}
\! \!\right)(x^\mu,y ) \! \!&=&\! \!
\left(\! \!\begin{array}{l}\, \,  \,\{\psi_{e}, \psi^\mu_{e}\},
- \{\psi^D_{o} \}
\\ - \{ \psi_{o}, \psi^\mu_{o}\}, \, \,  \,\{\psi^D_{e} \}\end{array}
\! \!\right)(x^\mu,- y).
\eea
This implies that the fermion $\Psi$ has the following Fourier
decomposition :
\bea
\psi_{e}(x^\mu,y) &=& \sum_{n=0}^\infty \psi^{(n)}_{e}(x^\mu)
\cos{(\frac{n+1/2}{R} y)}\cr
\psi_{o}(x^\mu,y) &=& \sum_{n=0}^\infty \psi^{(n)}_{o}(x^\mu)
\sin{(\frac{n+1/2}{R} y)}
\eea
and in a similar way for the gravitino components:
\bea
\psi^M_{e}(x^\mu,y) &=& \sum_{n=0}^\infty \psi^{M(n)}_{e}(x^\mu)
\cos{(\frac{n+1/2}{R} y)}\cr
\psi^M_{o}(x^\mu,y) &=& \sum_{n=0}^\infty \psi^{M(n)}_{o}(x^\mu)
\sin{(\frac{n+1/2}{R} y)}\, .
\eea
Note that there are no massless fermions in $D$-dimensional space, as all
Kaluza-Klein (KK) modes have mass-shift by half a unit.

Consider now the action of the $\mathbb{Z}_2$ orbifold  on the
supersymmetric transformations. The spinorial parameter $\eta$ of the
transformation is also decomposed in two parts, an even and an odd one:
\bea
\eta = \left(\begin{array}{l}\eta_{e}\\ \eta_{o}\end{array}
\right).
\eea
The bosonic fields are periodic and should remain as such after
supersymmetric transformations. As a result, the supersymmetry
transformation parameter should be anti-periodic and can be Fourier
decomposed as:
\bea
\eta_{e}(x^\mu,y) &=& \sum_{n=0}^\infty \eta^{(n)}_{e}(x^\mu)
\cos{(\frac{n+1/2}{R} y)}\cr
\eta_{o}(x^\mu,y) &=& \sum_{n=0}^\infty \eta^{(n)}_{o}(x^\mu)
\sin{(\frac{n+1/2}{R} y)}
\eea
A few important remarks follow from the above analysis:
\begin{itemize}
\item At $y=0$ the odd component of $\eta$ vanishes: $\eta_{o} =0$.
Half of the original supersymmetry transformations, associated with the
spinor $\eta_{e} (x^\mu, 0)\equiv \eta_{e} (x^\mu)$, remain. The
associated gravitino  $\psi^\mu_{e}$ survives and can have localized
supersymmetric coupling with matter fields at $y=0$. The other gravitino
$\psi^\mu_{o}$ has vanishing wave function and  can have only $y$-derivative
couplings at $y=0$. On the boundary at $y=\pi R$, the other half of
 supersymmetry associated with
$\eta_{o}(x^\mu, \pi R)\equiv \eta_{o}(x^\mu)$ is preserved.
We will denote by $Q_e$ and $Q_o$ the supersymmetry generators associated
with $\eta_{e} (x^\mu)$ and $\eta_{o}(x^\mu)$, respectively.

\item The fermion $\eta$ has no massless modes in $D$ dimensions. In
the Fourier expansion, all modes $\eta^{(n)}_{e}$ and $\eta^{(n)}_{o}$
are massive. Thus, in the $D$-dimensio\-nal theory, there is no
supersymmetric transformation leftover, i.e. supersymmetry is totally
broken.
\end{itemize}

\vspace{0.5cm}

\noindent {\bf  (ii) {\it  Appearance of orientifolds}}

\vspace{0.3cm}

The above discussion is generic to all string vacuua. We will now
consider the case of weakly coupled type IIB orientifolds \cite{tbo,reviews}. 
These contain
two types of extended objects, orientifold planes (O$_p$-planes) and
D$p$-branes, whose world-volume is extending in $p+1$ dimensions. They
carry two types  of charges Ramond--Ramond (RR) and
Neuveu--Schwarz--Neuveu--Schwarz  (NS--NS) charges as listed\footnote{
O$_p$ and O$^{\prime}_p$ are often denoted  in the literature as O$_p^-$ and
O$_p^+$, respectively.} in Table 1.
For the purpose of our discussion it is important to remind that:

\begin{table}[h]
\begin{center}
\renewcommand{\arraystretch}{0.3}
\begin{tabular}{  | c || c | c | c | c || c | l |}
\hline &  \mco{4}{c||}{}  &\mco{2}{c|}{}
\\ & \mco{4}{c||}{Orientifolds}&\mco{2}{c|}{D-branes}\\
   &  \mco{4}{c||}{}  &\mco{2}{c|}{}
\\ \hline\hline & & & & & & \\
   Symbol  & \, \, O$_p$ \, \, &  \, \,$\bO_p$  \, \, &  \, 
\,O$^{\prime}_p$ \, \,
   & \, \, $\bOp_p$ \, \, & \, \, D$p$ \, \, & \, \,  $\bD p$  \, \\
   & & & & & & \\
\hline  & & & & & & \\ RR charge  & $-$ &  + & + & $-$ & +& \, \, $-$ \\
   & & & & & & \\ \hline  & & & & & & \\
   NS--NS charge & $-$& $-$ &+ & + &+ & \,  \, + \\  & & & & & & \\
   \hline
\end{tabular}
\end{center}
\caption{The RR and NS--NS charges of orientifolds and D-branes.
}\end{table}

\begin{itemize}
\item The tension of these extended objects is proportional to their
NS--NS charge. This implies in particular that the orientifold O$_p$ and
$\bO_p$ planes have negative tensions and therefore they are not
dynamical objects, i.e. they  do not have massless  fluctuations.
Requiring vanishing tree-level cosmological constant amouts in our
construction to impose zero total NS--NS charge.

\item The total RR charge of the system in the compact internal space
should vanish by Gauss law (see  
tadpole cancellation conditions in section 4).

\item Each brane and orientifold plane breaks  by itself
half of the bulk supersymmetry. The conserved (broken) half is linearly
(non-linearly) realized on the world-volume of the D-brane.
  In our notations,
O$_p$, O$^{\prime}_p$ and D$p$ conserve the supersymmetries 
associated to $Q_e$,
while $\bO_p$, $\bOp_p$ and $\bD p$ conserve $Q_o$.
\end{itemize}

Since the orientifold planes act as mirrors, changing the orientation of
strings when going through them, they can appear only at the boundaries
of the compactification interval $S^1/Z_2$. We will consider here configurations where an
O$_{D-1}$ and an $\bO_{D-1}$ planes  are sitting at $y=0$ and $y=\pi R$,
respectively. Note, that these do not break any further the part of
supersymmetry leftover by the orbifold projection.

\vspace{0.5cm}

\noindent {\bf  (iii) {\it  Adding D-branes}}

\vspace{0.3cm}

We can now introduce D-branes at the boundaries. The previous choice
of orientifold planes carrying negative tension
allows to compensate the vacuum energy arising from the tensions of
the D-branes, in order to keep the tree-level cosmological constant
vanishing. To avoid appearance of a global RR charge we can only
add pairs of  D$p$--$\bD p$ branes. We will consider two possibilities
for the positions of these  branes:

\begin{itemize}
\item {\underline {Model I :}}  the D$p$-branes are put at
$y=0$ while the antibranes $\bD p$ are put at $y=\pi R$. This
ensures local cancellation of all tree-level tadpoles and leads to a
gauge theory with  linearly realized supersymmetries associated to $Q_e$
and $Q_o$ on the branes at $y= 0$ and $y=\pi R$, respectively.
   The Scherk-Schwarz boundary conditions allow supersymmetry
in the bulk to interpolate between $Q_e$
and $Q_o$  when going from one
boundary to the other. Such a brane model has been studied in detail in
refs. \cite{ads1,HF}.

\item {\underline {Model II :}} The other possibility is to put
the D$p$-branes on top of an $\bO_p$ at $y=\pi R$ and the
$\bD p$-branes at $y=0$. At each boundary, the branes and orientifolds
preserve  a different half of the original supersymmetry. Thus their
superposition breaks it totally both at $y=0$ and $y=\pi R$.  The
supersymmetries associated with $Q_e$ and  $Q_o$ are  non-linearly
realized on the  world volumes  of $\bD p$ and D$p$ branes,
respectively. In fact, it is possible to identifiy a gauge singlet fermion
on each boundary with the Goldstino and the  scale of supersymmetry
breaking corresponds to the one computed in section 2.
\begin{figure}[htb]
\centering
\epsfxsize=4.5in
\epsfysize=4.5in
\epsffile{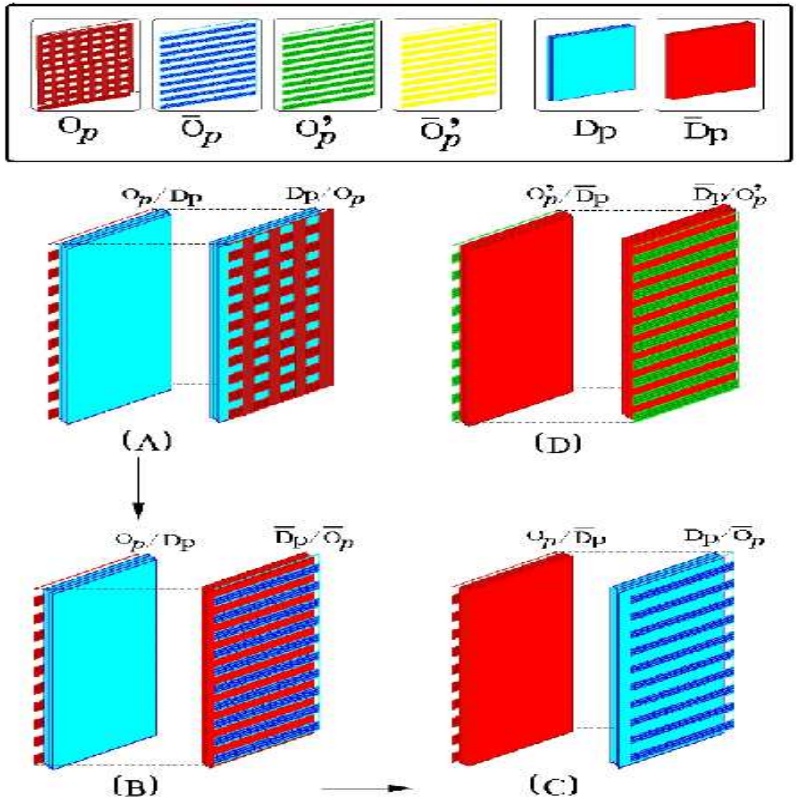}
\caption{\it (A) Supersymmetric configuration (B) a
$\mathbb{Z}_2$ Scherk-Schwarz deformation leads to supersymmetry breaking 
in the bulk
(C) Switching the positions of branes and anti-branes leads to
supersymmetry breaking also on the branes. The case (D) corresponding to
previously considered non-supersymmetric brane-worlds \cite{ads2,Sug}
is displayed for comparison. }
\end{figure}

\end{itemize}

This construction is summarized in figure 1. We will discuss explicit
realizations of this scenario in the following. Here, we would like to
comment about the origin of non-linear supersymmetry on the boundaries.
Consider first a stack of $N$ coincident (anti) D-branes in the bulk
(away from the boundaries) with a $U(N)$ Yang-Mills theory on their
world-volume. These are solitonic objects which break spontaneously half
of the bulk supersymmetry. Thus, the leftover supersymmetry is realized
linearly while the broken half becomes non-linear. The 
corresponding
Goldstino can be identified with the abelian gaugino of the $U(1)$ factor
in $U(N)$ \cite{Bagger}. When this collection of (anti) D-branes is put on 
top of an
(orientifold) anti-orientifold O-plane, say at the origin $y=0$, the
orientifold  projection breaks explicitly the linear supersymmetry on
the branes as it acts differently on fermions and bosons. However, the
non-linear supersymmetry on their world-volume is preserved.

Another aspect of the non-linear realization of global supersymmetry is
the presence of a vacuum energy. This is related to the Goldstino decay
constant which was computed in section 2 and (for D3-branes) is given
by $v^4/2$. When coupled to supergravity with vanishing cosmological
constant, the gravitino mass is given by
$m_{3/2}= v^2/M_{pl}$, where $M_{pl}$ is the four-dimensional Planck
mass \cite{DZ}. Note that this relation does not hold in our case, since there is
an additional source of supersymmetry breaking due to the Scherk-Schwarz
boundary conditions in the bulk and the (tree-level) gravitino
mass is given by $1/2R$. In fact, in our model the tree-level
contribution to the vacuum energy from the supersymmetry breaking is
zero as a result of cancellations of different contributions among
(positive tension) branes and (negative tension) orientifolds.


\begin{table}[h]
\begin{small}
\begin{center}
\renewcommand{\arraystretch}{1.3}
\begin{tabular}{  | c | c || c || c | c | c | c | c || c || c | c|}
\cline{1-5} \cline{7-11}
\mco{2}{|c||}{$y=0$}& Bulk &\mco{2}{c|}{$y=\pi R$}&  &
\mco{2}{c||}{$y=0$}& Bulk &\mco{2}{c|}{$y=\pi R$} \\
\cline{1-5} \cline{7-11}
DD & DN &{\small closed}/NN & $\bD $N   & $\bD \bD$ &  \,
&  $\bD \bD$ & $\bD $N &{\small closed}/NN & DN  & DD
\\ \cline{1-5} \cline{7-11}
  &  &\, $e^{M}_a$  &  &  & &  &  & $e^{M}_a$ &
  & \\ &  & $\psi^{\mu}_+$ \,  $\psi^{\mu}_-$ &  &  &
  &  &  &  $\psi^{\mu}_+$ \,  $\psi^{\mu}_-$ &
& \\ &  & $\psi^{D}_-$ \,  $\psi^{D}_+$&  &  &
  &  &  & $\psi^{D}_-$ \,  $\psi^{D}_+$  &
& \\
$A_\mu$ &  &\, $A^{(b)}_M$  &  & $A_\mu$ & & $A_\mu$ &  &$A^{(b)}_M$ &
  &$A_\mu$ \\
$\lambda_+$ &  & $\lambda^{(b)}_+$ \,  $\lambda^{(b)}_-$&  & $\lambda_-$ &
  & $\lambda_-$ &  &  $\lambda^{(b)}_+$ \,  $\lambda^{(b)}_-$ &
&$\lambda_+$ \\
$\psi_-$ &  &  $\psi^{(b)}_-$ \,  $\psi^{(b)}_+$&  &$\psi_+$ & & 
$\psi_+$ &  &  $\psi^{(b)}_-$ \,  $\psi^{(b)}_+$ &  &$\psi_-$ \\
$\phi$ & & $\phi^{(b)}$  &  &$\phi$ & & $\phi$ &  & $\phi^{(b)}$ &  &$\phi$ \\
  & $\psi'_-$ &  & $\psi'_+$ & & &  & $\psi'_-$ &  & $\psi'_+$ & \\
  & $\phi'$ &  & $\phi'$ & & &  & $\phi'$ &  & $\phi'$ & \\
\cline{1-2} \cline{4-5}\cline{7-8} \cline{10-11}\mco{2}{|c||}{Linear 
$Q_e$}& &\mco{2}{c|}{Linear $Q_o$}& & \mco{2}{c||}{Non-L $Q_e$}& 
&\mco{2}{c|}
{Non-L $Q_o$}\\
\cline{1-5} \cline{7-11} \mco{5}{|c|}{Model I}& & \mco{5}{c|}{Model II}\\
\cline{1-5} \cline{7-11}
\end{tabular}
\end{center}
\caption{Comparison of Model I and Model II. In the first, $Q_e$ and $Q_o$
supersymmetries are linearly realized at $y=0$ and $y=\pi R$, respectively.
In the second they are non-linearly realized.
  The labels + and $-$ represent
the chiralities, $\lambda_{\pm}$ are fermions in the adjoint representations,
gauginos in Model I. The label b stands for bulk states.
}
\end{small}
\end{table}


\subsection{Main features of the construction}

The first important remark we should make is about presence of
possible tachyons. The branes and anti-branes attract each other,
leading at short distances to the appearance of tachyonic modes for
open strings stretched between the two sets. To avoid them, it is
necessary that the dimension employed by the Scherk-Schwarz deformation
is  much larger than the string scale. This is fortunately the regime
where the effective field theory analysis carried above is valid.

Another important issue is stability of masses under radiative
corrections. At tree-level, the KK spectrum of bulk fields is the
same for both  models I and II. In the explicit examples given in section
4, this is  manifest from the fact that the torus and Klein bottle
amplitudes are the same for the two configurations. At one-loop, in model
II, new contributions to the gravitino and other masses are expected to
arise from radiative corrections of the boundary states that are not
supersymmetric. Naively, these are of the order of $M_s^2/M_{pl}$,
although a careful analysis is needed to include the effects of all KK
excitations, as well as the contributions of both boundaries. Note that
$M_s^2/M_{pl}$ is of the same order as the tree-level masses given by the
compactification scale $1/R$ only for the case of two transverse
dimensions. For more than two, $1/R$ is dominant, while for one
$M_s^2/M_{pl}$ is larger. However, the case of bulk propagation in
one dimension is subtle because of the expected large corrections growing
linearly with the radius, originating from local tadpoles \cite{AB}.
A similar question concerns the value of the Goldstino mass due to
radiative corrections from the bulk, where local supersymmetry is
spontaneously broken by the Scherk-Schwarz boundary conditions.
We plan to return to this issues in a future work.

Let us also discuss some phenomenological applications of these models.
If the Standard Model is localized on the world-volume of
non-supersymmetric D-branes considered here, then the gauge
hierarchy problem requires that the string scale should lie in the TeV
region. It is then preferable to choose the extra-dimension separating
the brane--antibrane sets to be part of the gravitational bulk with a
size as large as a millimeter \cite{add}.

In order to evade the above constraint  on the string scale, we need to
embed the Standard Model gauge group on other branes with
supersymmetric massless spectrum. Examples of such configurations are
illustrated in figure 2. The non-supersymmetric branes act then as
``hidden sectors'' and bulk fields mediate the supersymmetry breaking to
the Standard Model branes. The study of the issues of radiative
corrections discussed above is now necessary to fix the desired values
for the string and compactification scales \cite{intermediate}.

\begin{figure}[htb]
\centering
\epsfxsize=5.5in
\epsfysize=3.0in
\epsffile{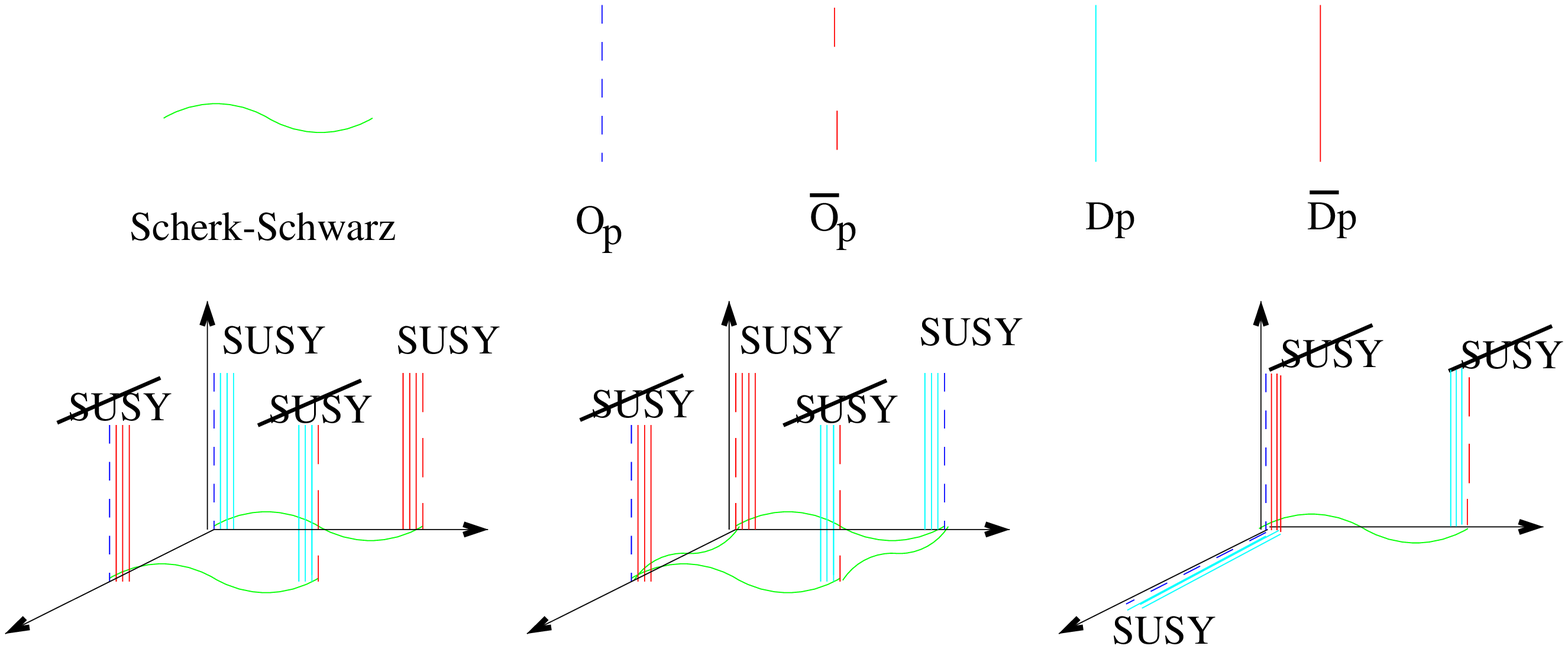}
\caption{\it  Constructions containing `` tree-level supersymmetric 
sectors''. The configurations with branes on top of orientifolds with 
the same RR charge are non-supersymmetric while those with opposite RR 
charges lead to supersymmetric massless spectrum. In the first figure, the 
Scherk-Schwarz boundary condition is on one direction, while in the second 
figure, it is imposed in two dirrections. }
\end{figure}

\subsection{ Summary of the explicit examples}

In the next section we will present explicit realizations of the 
construction described here. We will first consider the simplest case 
with a single  compact dimension $S^1/{\mathbb Z}_2$ and only one type of
branes and orientifolds, D8 and O$_8$. We will show that in the presence of a 
Scherk-Schwarz breaking of supersymmetry in the bulk, the tadpole 
cancellation conditions have two solutions. In the first solution, 
the boundary 
massless states form supersymmetric multiplets while in the second they
don't. More precisely, on each of the boundaries lives an $SO(16)$ Yang-Mills 
theory. In the first model, the fermions are in the anti-symmetric 
representation of $SO(16)$ and form  vector supermultiplets with the gauge 
bosons, while in the second the fermions are in the symmetric representations 
and do not 
have massless supersymmetric partners. In order to illustrate that these 
constructions realize model I and II described above, we show that
one can go from one solution to the other by interchanging the positions of 
 branes on the two boundaries. We will also
show  that the RR charges at the boundaries vanish in the first solution 
while they have opposite values in the second. This realizes the above 
description in terms of an O$_8$--$\bO_8$ system on which there are stacked 
D$8$--$\bD 8$ branes.
In both cases, the bulk states contain only the ten-dimensional supergravity 
multiplet where the fermions, gravitino and dilatino have mass shifts by $1/2R$.

A simple toroidal compactification of the  above nine-dimensional 
example does not lead to four-dimensional
chiral fermions. 
To obtain chiral four-dimensional models, we need to 
perform further orbifolding. A simple example consider here 
is given by  
$T^6 / ({\mathbb {Z}_2 }
\times {\mathbb {Z}_3 }) \equiv T^6 / {\mathbb {Z}^{\prime}_6 }$ 
and will be constructed in two steps: first we  obtain a chiral 
six-dimensional model through compactification on $T^4 / {\mathbb {Z}_2 }$, 
and second we perform a further compactification on $T^2$ followed by 
a ${\mathbb {Z}_3 }$ projection to obtain a chiral four-dimensional theory.

The six-dimensional example obtained by compactification on 
$T^4 / {\mathbb {Z}_2 }$ with a Scherk-Schwarz breaking of supersymmetry 
along one of the $T^4$ compact directions has some interesting features.
First, it contains two types of branes D5 and D9. The D9 branes 
extend in the whole 
space-time and give rise to Yang-Mills theory coupled with matter in the bulk 
while the D5's give rise to  chiral  gauge theories localized on the 
boundaries. It provides then an example with the most generic bulk+boundaries 
content as described in Table 2. In the bulk, on one hand the 
closed string modes lead
to massless graviton,  anti-symmetric tensor,  dilaton and
sixteen  scalars. On the other hand, the  open string modes
from D9 branes give rise to the gauge bosons of $U(16)$ and four scalars 
in the 
${\bf 120 + \ov{120}}$ representations.  
The fermionic supersymmetric partners of these bulk states are massive 
because of the Scherk-Schwarz boundary conditions.
On the boundaries, the  closed string
twisted states  form supersymmetric
multiplets. These represent sixteen neutral hypermultiplets
localized at each of the sixteen $T^4/{\mathbb{Z}_2}$ fixed points;
half of them located at $y=0$ have negative chirality, while the other
half located at $y= \pi { R}$  have positive chirality. The open 
string states lead to $U(8)$ gauge theories on the boundaries.
The matter content of Model II is then listed in Table 3. It differs from 
the spectrum of Model I first in the chirality assignments as explained in Table 2, and
second because the fermions are in symmetric representations in 
Model I and in anti-symmetric ones in Model II. The symmetric representation $36$ of 
$U(8)$ can in fact be decomposed into the irreducible components $35 +1$,
where the singlet is 
identified with the Goldstino of the supersymmetry which becomes 
non-linearly realized  in Model II\footnote{More precisely, in the presence of 59 strings, the Goldstino is a linear combination of 99 and 55 states, as we described in section 2.}. 

From the six-dimensional example, it is easy to construct a chiral 
four-dimensional descendant on $T^6 / ({\mathbb {Z}_2 }
\times {\mathbb {Z}_3 })$ orbifold. The ${\mathbb {Z}_3 }$ orbifolding
 does not lead to new open string states but acts only as a projection on 
the six-dimensional spectrum. The resulting open string massless modes 
are listed in table 4. The four-dimensional 
Goldstino of the non-linearly realized 
supersymmetry is also easily identified with the ${\mathbb {Z}_3 }$-invariant
component of the corresponding six-dimensional spinor. It is interesting to 
note 
that without much efforts, we obtain in this way a two-family version of 
Pati-Salam model localized on the boundaries.

The construction  presented here can be modified in several ways.
  The models we describe have a symmetry between theories living on the
two boundaries. There are mirror worlds related by a chirality flip.
However, as  in the Ho\v rava--Witten compactification of M-theory,
  this symmetry could be broken upon  compactification.

\section{Non-supersymmetric D8-branes}

The construction of the nine-dimensional model that we describe here as
an orientifold from type IIB is quite simple and was given in ref.
\cite{ads1}. The generic type I string partition function is given as
the sum of four contributions corresponding to the diagrams represented
by the torus, the Klein bottle, the annulus and the M{\"o}bius worldsheet
surfaces \cite{tbo}. The corresponding amplitudes are denoted by the symbols
${\bf T}$, ${\bf K}$, ${\bf A}$ and ${\bf M}$, respectively. The general
expressions for these amplitudes together with our notations are given in 
the appendices A and B.

The simplest example of string models with non-linear brane supersymmetry
and vanishing tree-level vacuum energy is obtained as an orientifold of
type IIB string compactified on $S^1$ of radius $R$. It can also be
described in its $T$-dual version as a compactification of the
ten-dimensional type IIA string on $S^1/\mathbb{Z}_2$ of radius
${\tilde R} = 1/R$. Below, we will often make use of the type IIA
description which has natural geometrical interpretation in the region
${\tilde R}>l_s$ in terms of D8-branes, while the partition function will
be given for convenience in terms of type IIB variables.

We consider a compactification of type IIA string on an orbifold
$S^1/\mathbb{Z}_2$ parametrized by the coordinate $y \in [0,\pi {\tilde
R}]$.  We will require the presence of~:
\begin{itemize}
\item {\it At $y=0$:} $n_{0\epsilon}$ $D{8}$-branes, each carrying a RR
charge $\epsilon= \pm 1$, stacked on an $O_8$ orientifold plane.

\item {\it At $y=\pi {\tilde R}$:} $n_{\pi \epsilon}$ of
$D{8}$-branes, each carrying  a RR charge $-\epsilon$,
stacked on an ${\ov O}_8$ orientifold plane.

\item {\it Along}~ $y \in [0,\pi {\tilde R}]$: a Scherk-Schwarz
deformation shifting all bulk fermions masses by $1/2R$.
\end{itemize}

The perturbative spectrum of the model can be read off from the one-loop
partition function. Using type IIB variables, it can be written as:
\bea
\frac{1}{ 2}{\bf T}+ {\bf K}+{\bf A}+{\bf M}=
\int\left(\frac {1}{ 2}{\cal T}(\tau,\bar{\tau})+{\cal K}(2i\tau_2)
+{\cal A}(\frac{it}{2})+{\cal M}(\frac{it}{2}+\frac{1}{2})\right)
\label{typeI}
\eea
where the integration measure for
torus, Klein bottle, annulus and M{\"o}bius contributions
  are defined by:
\bea
{\bf T}&=&\frac{1}{({4 \pi^2 {\alpha'}})^{\frac{9}{2}}}
\int \frac {d \tau d {\bar \tau}}{({\rm Im} \ \tau)^{11/2}} \, \, 
| \frac {1}{\eta^7}|^2\, \, 
{\cal T} \ ,\nonumber \\
{\bf K}&=&\frac{1}{ ({4 \pi^2 {\alpha'}})^{9/2}}
\ \int_0^{\infty} \frac{d \tau_2}{\tau_2^{11/2}}\, \, 
\frac{1}{\eta^7} \, \, {\cal K}
, \nonumber \\
{\bf A}&=& \ \ \frac{1}{({8 \pi^2 {\alpha'}})^{9/2}}
\int_0^{\infty} \frac{d t}{t^{11/2}} \, \,  \frac{1}{\eta^7}
\, \, {\cal A} 
\nonumber \\
{\bf M}&=& \  \ {1 \over ({8 \pi^2 {\alpha'}})^{9/2}}
\int_0^{\infty} {d t \over t^{11/2}} \, \,  {1 \over \eta^7}\, \,  {\cal M}
.  \label{amnnine}
\eea
The integrands ${\cal T}$, ${\cal K}$,
${\cal A}$ and ${\cal M}$ can be written in a compact way as:
\bea
{\cal T}\!\!\!&=&\!\!\!
  E_0\, \,(V_8\, \, \overline{V_8}+S_8\, \, \overline{S_8})
+O_0\, \,(I_8\, \, \overline{I_8}+C_8 \, \, \overline{C_8})
-E_{\frac{1}{2}}\, \,(V_8\, \, \overline{S_8}+S_8\, \, \overline{V_8})
\nonumber \\ &-& O_{\frac{1}{2}}\, \,(I_8\, \, \overline{C_8}
+C_8\, \, \overline{I_8}) \nonumber \\ \nonumber \\
{\cal K}\!\!\! &=&\frac{1}{2}(V_8-S_8)\, \,\sum_m Z_{2m}+\frac{1}{2}
(I_8-C_8)\, \,\sum_m Z_{2m+1}\label{part} \\ \nonumber \\
{\cal A}\!\!\! &=&\!\!\!\! \left[\!
\frac{n_{0+}^2\!+n_{\pi +}^2 \!+ n_{0-}^2\!
+n_{\pi -}^2}{2}(V_8-S_8)
+(n_{0+} n_{0-}\!+\! n_{\pi +}n_{\pi -})(I_8-C_8)
  \right]
\!\sum_m Z_{2m}
  \nonumber \\ &+&\!\! \left[
(n_{0+}n_{\pi -}\!+\!n_{0-}n_{\pi +})(V_8-S_8)
\!+\!(n_{0+}n_{\pi +}\!+\! n_{0-}n_{\pi -})(I_8-C_8)
\right]\sum_m Z_{2(m+\frac{1}{2})}
\nonumber \\ \nonumber \\
{\cal M}\!\!\! &=& \sum_m \left[-\frac{n_{0+}+n_{\pi 
+}}{2}({\hat V_8}-(-1)^m {\hat S}_8)\!-\! \frac{n_{0-}+n_{\pi 
-}}{2}({\hat V_8}+(-1)^m {\hat S}_8)
\right] Z_{2m}\nonumber
\eea
in terms of $SO(8)$ characters and lattice sums as defined in the
appendix A.

The cancellation of global tadpoles (vanishing of the total NS-NS and RR
charges) requires the coefficients of the massless modes contributing to
the characters $V_8$ and $S_8$ in the transverse closed string channel
 to vanish. This implies:
\bea
(n_{0+}\!+\!n_{\pi +})\!+\!(n_{0-}\!+\!n_{\pi -})= 32
\ ,\qquad
(n_{0+}\!-\!n_{\pi +})\!-\!(n_{0-}\!-\!n_{\pi -})=  0
\label{tad9}
\eea
Only the annulus and M{\"o}bius are sensitive to the RR charges of the
branes. For instance, the torus amplitude in eq. (\ref{part}) is
directly obtained from the supersymmetric expression by a simple
deformation corresponding to shifting the masses of all fermionic KK
modes of closed strings, due to the Sherck-Schwarz breaking along
the tenth coordinate, and complete the partition function in the
non-trivial winding sector using modular invariance \cite{Kounnas}.

The simultaneous presence of branes and anti-branes leads to instability
that manifests in  the appearance of tachyons associated to the identity
character $I_8$ in eq. (\ref{part}). All these tachyonic modes acquire
positive mass-squared in the limit
$R \rightarrow 0$ (${\tilde R} \rightarrow \infty$) except for
the ones appearing in the sector $I_8 Z_{2m}$ for $m=0$. To remove this
tachyon we choose  $n_{0\epsilon}= n_{\pi \epsilon} =0$ for either
$\epsilon =+$ or $\epsilon =-$. This condition amounts to require that
only branes (or only anti-branes) are present at $y=0$, while only
anti-branes (or only branes) carrying the opposite RR charge are present
at $y=\pi R$. As a result, there are two possible choices:
\bea
{\rm Model~ I:}&& n_{0+}= n_{\pi +} =16,
\qquad  n_{0-}= n_{\pi -} =0 \\
{\rm Model~ II:}&& n_{0+}= n_{\pi +} =0,
\qquad \, \, \,  n_{0-}= n_{\pi -} =16
\eea

In Model I, studied in great detail in ref. \cite{ads1}, supersymmetry is
broken at tree-level only in the bulk, while on the boundaries the
massless states form supersymmetric multiplets: a vector and a spinor in
the adjoint of $SO(16) \otimes SO(16)$. On one boundary, D8-branes are
stacked on top of an $O_8$ plane, while on the other boundary,
$\bD 8$ branes sit on top of an ${\ov O}_8$. Note however that open string
winding modes are not supersymmetric, since they are obviously sensible
in global effects. In particular, fermions with odd winding numbers in
each of the two boundaries transform in the symmetric representation of
the gauge group, while bosons remain in the adjoint.

In contrast to Model I, in Model II, even the massless states on the
boundaries are non-supersymmetric: they form vectors in the adjoint
representation $(1,120) +(120,1)$ with a spinor in the symmetric
representation
$(135,1) +(1,1)$ and $(1,135) +(1,1)$ of $SO(16) \otimes SO(16)$.
The model corresponds to ${\ov {\rm D}}8$-branes sitting on top of $O_8$
on one boundary, and D8-branes sitting on top of ${\ov O}_8$ on the other
side. The gauge singlet fermions are identified with  the two localized
Goldstinos in each boundary, while the corresponding decay constant
can be computed as explained in section 2. It is given by $16 \tau_8$,
with $\tau_8=\frac{1}{(2\pi)^7}\frac{M_s^9}{g_s}$ the D8-brane tension. Note that in this case, odd
winding modes are supersymmetric.

\subsection{Interpolating between the two models}

In this section we will provide further evidence to the geometrical
interpretation of the models presented above. We will first show that one
can obtain  model I from model II, and vice-versa, by interchanging the
branes between the  two boundaries.  We will then compute the NS-NS and
RR charges  to verify the presence of orientifolds and branes with
opposite charges in the two ends, as described in section 3.

For this purpose, it is useful to
consider the generic situation where the D8 (or $\bD 8$)
  branes are not sitting at the orbifold fixed points.
Half of the branes are moved away from the origin $y=0$, at
a distance
$2 \pi {\tilde R} {\vec a}= ( 2 \pi {\tilde R}a_1,\cdots,
2 \pi {\tilde R}a_{16}) $, while the other half
are away from  $y= \pi {\tilde R}$ by
$2 \pi {\tilde R}{\vec a'}= (2 \pi {\tilde R}a'_1 ,\cdots,
  2 \pi {\tilde R}a'_{16})$.

In the type IIB representation, the brane separations are
described by Wilson lines appearing as shifts in the KK momenta of open
strings. These contribute to the annulus and M{\"o}bius amplitudes
which become:
\bea
{\cal A}&=&\!\!\!\frac{1}{2}(V_8-S_8)
\left[\sum_{i,j,m}Z_{2(m+a_i+a_j)}+\sum_{i,j,m}
Z_{2(m+a'_i+a'_j)}\right] \\
&&+(I_8-C_8)\sum_{i,j,m}Z_{2(m+\frac{1}{2}+a_i+a'_j)}\nonumber 
\\
-2 {\cal M}&=&\!\!\!\sum_{i,m}Z_{2(m+2a_i)}
(\hat{V}_8-(-1)^m\hat{S}_8)+\sum_{i,m}Z_{2(m+2a'_i)}(\hat{V}_8-(-1)^m\hat{S}_8)\nonumber
\eea
  As the torus and  Klein bottle  amplitudes
(in the direct channel) get contributions only from closed strings,
  they are not affected. The two cases of interest are:
\begin{itemize}
\item {\underline {${\bf\vec a} = {\bf\vec a'} ={\vec 0}$}} which
leads to:
\bea
{\cal A}&=&{2^8}(V_8-S_8)  \sum_{m}Z_{2m}
+2^8 (I_8-C_8)  \sum_{m}Z_{2(m+\frac{1}{2})}\nonumber \\
  {\cal M}&=&\!\!\!   -2^4 \sum_{m}Z_{2m}
(\hat{V}_8-(-1)^m\hat{S}_8)
\eea
and reproduces the corresponding spectrum of Model I.

\item{\underline {${\bf\vec a} = {\bf\vec a'} ={\vec \frac{1}{2}}$}}
  which leads instead to:
\bea
{\cal A}&=&{2^8}(V_8-S_8)  \sum_{m}Z_{2m}
+2^8 (I_8-C_8)  \sum_{m}Z_{2(m+\frac{1}{2})}\nonumber \\
  {\cal M}&=&\!\!\!   -2^4 \sum_{m}Z_{2m}
(\hat{V}_8+(-1)^m\hat{S}_8)
\eea
describing Model II.
\end{itemize}
This shows that the two models are related through a shift by $\pi
{\tilde R}$ in the positions (i.e. an interchange) of the D-branes.

In the expression of the transverse channel for the one-loop amplitudes,
the coefficients of the characters  $V_8$ and $S_8$ give
the local NS-NS and R-R tadpoles, respectively. Summing up the
Klein, annulus and  M{\"o}bius contributions  we obtain:
\bea
&&\!\!\!\! \frac{R}{2^{\frac{3}{2}}}V_8\sum_n \left[
\{32-2(trW^{2n}+tr{W'}^{2n})\}\tilde{Z}_{2n}
+\frac{1}{32}(trW^n+(-1)^ntr{W'}^n)^2\tilde{Z}_n
\right]
\nonumber
\\
- \frac{R}{2^{\frac{3}{2}}}\!\!\!\!&&\!\!\!\!S_8\sum_n \left[
\{32 +2(trW^{2n+1}\!+\!tr{W'}^{2n+1})\}\tilde{Z}_{2n+1}
\!+\!\frac{1}{32}(trW^n\!\!-\!(-1)^ntr{W'}^n)^2\!\tilde{Z}_n
\right]\nonumber
\eea
where the phases
\bea
W &=&
diag(e^{2i\pi\, a_1},\cdots,e^{2i\pi\, a_{16}})\nonumber \\
  W' &=&
diag(e^{2i\pi\, a'_1},\cdots,e^{2i\pi\, a'_{16}})
\eea
contain the effects of displacement of the branes from the boundaries,
so that $W=W'=1$ reproduces Model I while $W=W'=-1$ is
associated with Model II.

After some straightforward manipulations, we can express the NS-NS
tadpole as~:
\bea
\frac{R }{2^{\frac{13}{2}}}V_8 \sum_n\{\left[tr(2-W^{2n}-{W'}^{2n})
\right]^2\tilde{Z}_{2n}+\left[tr(W^{2n+1}-{W'}^{2n+1})\right]^2
\tilde{Z}_{2n+1}\}
\eea
which cancels for both $W=W'=1$ and $W=W'=-1$. On the other hand,
the RR tadpole is given by:
\bea
-\frac{R }{2^{\frac{13}{2}}}S_8 \sum_n\{\left[tr(2-W^{2n+1}-
{W'}^{2n+1})
\right]^2\tilde{Z}_{2n+1}+\left[tr(W^{2n}-{W'}^{2n})\right]^2
\tilde{Z}_{2n}\}
\eea
which vanishes for $W=W'=1$ but not for $W=W'=-1$, leading:
\bea
-\frac{R}{2^{\frac{3}{2}}}S_8~ (4\times32\sum_n\tilde{Z}_{2n+1})\, .
\label{tadRR}
\eea
It follows that while in model I branes are sitting on top of
orientifolds with  opposite RR charges, in Model II the RR charge of the
branes and  orientifolds are the same.

In eq. (\ref{tadRR}), only RR
states with  odd KK momenta feel the presence of charges which implies
that  the charges at the boundaries are opposite. In fact, let us denote
by $S$ the corresponding RR field. It can be Fourier expanded as:
\bea
S(x^\mu, y) = \sum_{n=0}^\infty S^{(n)}_{+}(x^\mu)
\cos{(\frac{n}{{\tilde R}} y)} + \sum_{n=1}^\infty S^{(n)}_{-}(x^\mu)
\sin{(\frac{n}{{\tilde R}} y)}
\eea
The coupling of the RR field to charges $q_0$ at $y=0$ and
$q_{\pi}$ at $y=\pi {\tilde R}$ can be written as:
\bea
\int d^{10}x \,  \L_{RR}\!\! &=& \!\!\int d^9x \int dy \{ \delta (y) 
q_0 S(x^\mu, y) + \delta (y-\pi {\tilde R})
q_{\pi} S(x^\mu, y)\} \nonumber \\
  &=& \!\! \int d^9x  \sum_{n=0}^\infty \left[ S^{(2n)}_{+}(x^\mu)
(q_0 + q_{\pi})
+ S^{(2n+1)}_{+}(x^\mu) (q_0 - q_{\pi}) \right]\, . \nonumber \\
\eea
Thus, the absence of ${\tilde Z}_{2n}$ in (\ref{tadRR}) corresponds to
$q_0 = - q_{\pi}$.

\section{Chiral six-dimensional example }

Our second example is a  six-dimensional type IIB model with D5, $\bD 5$
and D9 branes. Compared to the nine-dimensional example, two new features
appear here: the chirality of Majorana-Weyl spinors in six dimensions
allows to have smaller number of supersymmetry charges, and Yang-Mills
fields are now present on both boundaries and in the bulk.

The model is obtained  through compactification of type IIB
on a  $T^4/ {\mathbb{Z}}_2$ orbifold.
The ${\mathbb{Z}}_2$ action on the four coordinates
$x^6,x^7,x^8$ and $x^9 \equiv y$ compactified on circles of radii
$R^1, R^2, R^3$ and $R^4 =R$, respectively, is given by:
\bea
{x^6,x^7,x^8,y} \rightarrow {-x^6,-x^7,-x^8,-y}
\eea
and leads to $2^4$ fixed points where orientifolds are localized.
In the compact direction $y$, all bulk fermions $\Psi$ are chosen to
satisfy anti-periodic boundary conditions:
\bea
\Psi(x^\mu, y+ 2\pi R)= - \Psi (x^\mu, y)
\eea
which lead to a Scherk-Schwarz breaking of supersymmetry in six
dimensions \cite{ads1}.

The knowledge of the one-loop torus, Klein bottle, annulus and M{\"o}bius
amplitudes:
\bea
{\bf T}&=&{1 \over ({4 \pi^2 {\alpha'}})^{3}}
\int \frac{ d \tau d {\bar \tau}}{({\rm Im} \ \tau)^{4}}
\, \,  | {1 \over \eta^4}|^2
\, \, {\cal T} \ ,\\
{\bf K}&=&{1 \over  ({4 \pi^2 {\alpha'}})^{3}}
\ \int_0^{\infty} {d \tau_2 \over \tau_2^{4}}  {1 \over \eta^4} {\cal K}
, \nonumber \\
{\bf A}&=& \ \ {1 \over ({8 \pi^2 {\alpha'}})^{3}}
\int_0^{\infty} {d t \over t^{4}}  {1 \over \eta^4} {\cal A} 
, \nonumber \\
{\bf M}&=& \  \ {1 \over ({8 \pi^2 {\alpha'}})^{3}}
\int_0^{\infty} {d t \over t^{4}}  {1 \over \eta^4} {\cal M}
,  \label{amsix}
\eea
allows to derive the spectrum of the model.
In the model under study, the torus and Klein bottle amplitudes are the
same as in the model of ref. \cite{ads1} by construction, as we do not
affect the bulk states by switching the positions of the D5 and $\bD 5$
branes.

The torus amplitude is given by ${\cal T}={\cal T}_U+{\cal T}_T$ with:
\bea
{\cal T}_U \!\!&=& \!\!{\Lambda^{(3,3)} \over 2}
\left[ E^{\prime}_0 \, (\, |V_8|^2  + \,|S_8|^2  )
\!+\!  O^{\prime}_0 \, ( \,|I_8|^2  + \,
|C_8|^2 ) \right. \nonumber \\
  \!&&\!\left. \qquad \, - E^{\prime}_{1/2}\, ( V_8 \, {\bar S}_8 +\, 
S_8 {\bar V}_8 )
-\, O^{\prime}_{1/2} \, ( I_8\, {\bar C}_8 + \,C_8 \, {\bar I}_8 )
\right] \nonumber \\
{\cal T}_T \!\!&=&\!\!\! \frac {1}{ 4} (|Q_O\!-\!Q_V|^2+ |Q'_O\!-\!Q'_V|^2)
|4 {\eta^2 \over \theta_2^2}|^2 \
\!\! +\! \frac {1}{4}(|Q_S\!+\!Q_C|^2  \!+\!
|Q'_S\!+\!Q'_C|^2)
|4 {\eta^2 \over \theta_4^2}|^2 \nonumber \\
&&+
\frac {1}{4} (|Q_S\!-\!Q_C|^2\!+\!
|Q'_S\!-\!Q'_C|^2)
|4 {\eta^2 \over \theta_3^2}|^2
\nonumber
\eea
while the Klein bottle amplitude in the
direct channel is ${\cal K}= {\cal K}_U+ {\cal K}_T$ with:
\bea
{\cal K}_U&=&{1 \over 4} \left [ (V_8\!-\!S_8) (\sum_m Z_m \Lambda^{(3)}
\!+\!\sum_n {\tilde Z}_{2n}
{\tilde
\Lambda}^{(3)})\!+\!(I_8\!-\!C_8)\sum_n{\tilde Z}_{2n+1}{\tilde
\Lambda}^{(3)}\right]\nonumber \\
{\cal K}_T&=& (Q_S\!+\!Q_C\!+\!Q'_S\!+\!Q'_C)
{\eta^2 \over \theta_4^2}\,\,\,\,\,\,\,\,\,\,\,\,\,\,\,\,\,\,\,\,\,
\eea
Here 
${\cal T}_U$ and ${\cal K}_U$ represent the contributions of untwisted
closed strings, while ${\cal T}_T$ and ${\cal K}_T$ represent
the corresponding twisted closed string sectors. All the untwisted
massless fermions aquire masses $1/2R$ due to the Scherk-Schwarz
boundary conditions. Thus, the (bosonic) massless untwisted closed string
states contain  the graviton, the anti-symmetric tensor, the dilaton and
sixteen additional scalars. The twisted sector does not feel the
Scherk-Schwarz breaking  at tree-level and form supersymmetric
multiplets. These represent sixteen neutral hypermultiplets
localized at each of the sixteen $T^4/{\mathbb{Z}_2}$ fixed points;
half of them located at $y=0$ have negative chirality, while the other
half located at $y= \pi R$  have positive chirality. The change
of chirality arises from the Scherk-Schwarz boundary
conditions, which change half of the orientifolds O$_5$-planes
to $\bO_5$-planes as we explained in section 3.

The absence of tachyons is insured by separating
the branes from the anti-branes, thus taking the limit $R M_s >> 1$
\footnote{To be contrasted with the previous section, where in the type
IIB language $R M_s << 1$, here we have D5-branes of IIB, while before we
had D8-branes of IIA.}. We consider the Model II described in section 3,
obtained by putting $\bD 5$-branes  on top of the O$_5$ at $y=0$ and an
equal number of D5-branes  on top of  $\bO_5$ at $y= \pi  R$. The
one-loop amplitudes sensitive to the resulting breaking of
supersymmetry  on the boundaries are the annulus and M{\"o}bius
strip. The first is given in the direct (open string) channel by:
\bea
{\cal A}&=&
{n_N^2\over 4} (V_8\sum_mZ_{2m}\!-\!S_8\sum_mZ_{2(m+1/2)} )
\Lambda^{(3)}\nonumber \\
&+&\!
{1\over 4}\left[
(n_{D_1}^2\!+\!n_{D_2}^2)(V_8\!-\!S_8)\sum_n{\tilde Z}_{2n}
\!+\! 2 n_{D_1}n_{D_2}
(\! I_8\!-\!C_8)
\sum_n{\tilde Z}_{2(n+1/2)} \right]  {\tilde \Lambda}^{(3)}_0 \nonumber \\ 
\!\!&+&\!\!
{1\over 4}\left[
2n_Nn_{D_1}(Q_S+Q_C)+ 2n_Nn_{D_2}(Q'_S+Q'_C) \right]{\eta^2 \over \theta_4^2}
\nonumber \\
\!\!\!\!&+&\!\!\!\! \left[{1 \over 2} R_N^2 (Q_O-Q_V+Q'_O-Q'_V)+R_{D_2}^2
(Q_O-Q_V)+R_{D_1}^2
(Q'_O-Q'_V) \right] {\eta^2 \over \theta_2^2}  \nonumber \\
\!\!&+&\!\! {1\over 4}\left[
2R_NR_{D_2}(Q'_S-Q'_C)+ 2R_NR_{D_1}(Q_S-Q_C)
\right]{\eta^2 \over \theta_3^2}
\label{ann6}
\eea
while the M{\"o}bius amplitude  in the direct channel reads:
\bea
{\cal M} \!&=&\! - {n_N\over 4} ({\hat V}_8 \sum_m Z_{2m}\!-\!{\hat S}_8 
\sum_m Z_{2(m+1/2)})
  \Lambda^{(3)}\nonumber \\
&-& {n_{D_1}\!+n_{D_2}\over 4}
  \sum_n({\hat V}_8 {\tilde Z}_{2n}\!+\! \!{\hat S}_8 (-1)^n  {\tilde
Z}_{2n}) {\tilde
\Lambda}^{(3)}_0  \nonumber \\
&+&  \left[ n_N ({\hat V}_4{\hat I}_4\!-\!{\hat I}_4 {\hat V}_4)+n_{D_1}({\hat
Q}_O\!-\!{\hat Q}_V) +n_{D_2}({\hat Q'}_O\!-\!{\hat Q'}_V) \biggr)
  \frac { {\hat \eta}^2 } {{\hat \theta}_2^2}\right]
\label{S14}
\eea
where ${\Lambda}^{(3)}_0$ denotes the origin of the Narain's lattice.

While the effect of the Scherk-Schwarz breaking of supersymmetry in
the bulk was to transform the chirality of half of the twisted states
into the opposite chirality through $Q_{S,C} \rightarrow Q'_{S,C}$
\cite{ads1},  the effect of switching the position of the branes with that
of the anti-branes corresponds, first, to a permutation $R_{D_1}
\leftrightarrow R_{D_2}$ and $n_{D_1}\leftrightarrow n_{D_2}$, and
second, to shift $n$ to $n+1$ in ${\tilde Z}_{2n}$ of the M{\"o}bius
amplitude, which changes the sign of ${\hat S}_8 (-1)^n{\tilde Z}_{2n}$.

The tadpole cancellation conditions can be obtained from the sum of the
Klein bottle, annulus and M{\"o}bius  amplitudes expressed in the
transverse channel (see  appendix A).  There are two types of RR and
NS--NS fields, the twisted and the untwisted ones. From the former we get:
\bea
  R_N \ = \ R_{D_1} \ = \ R_{D_2} \ = \ 0 \, ,
\label{twtad}
\eea
while from the untwisted tadpoles we obtain:
\bea
  n_N = 32 \ , \qquad n_{D_1} =16 \,  \qquad
n_{D_2} = 16 \ .
\label{untwtad}
\eea
The conditions (\ref{twtad}) imply that the gauge groups are unitary and
their Chan-Paton charge spaces can be described through ``complex''
charges:
\bea
& &n_N = n + {\bar n} \ ,\qquad\qquad\quad
R_N = i(n - {\bar n}) \ ,\nonumber \\ & &n_{D_1} = m_1 + {\bar m}_1 \
,\qquad\quad\  n_{D_2} = m_2 + {\bar m}_2 \ ,\nonumber \\   & &R_{D_1}
= i(m_1 - {\bar m}_1) \ , \qquad R_{D_2} = i(m_2 - {\bar m}_2) \ .
\label{compcharges}
\eea
The tadpole cancellation conditions (\ref{untwtad}) lead then to:
\bea
n=16,\qquad m_1=8,\qquad m_2=8\, .
\eea

The massless open string spectrum can be read from the one-loop
amplitude:
\bea
{\A}_0+{\M}_0&=&(n{\bar n}+m_1{\bar m_1}+m_2{\bar m}_2)V_4I_4\
\nonumber
\\ &&+(\frac{n(n-1)}{2}+\frac{{\bar n}({\bar n}-1)}{2}+
\frac{m_1(m_1-1)}{2}\nonumber
\\ &&+\frac{{\bar m}_1({\bar m}_1-1)}{2}+
\frac{m_2(m_2-1)}{2}+\frac{{\bar m}_2
({\bar m}_2-1)}{2})I_4V_4\nonumber
\\ && -(\frac{m_1(m_1+1)}{2}+\frac{{\bar m}_1
({\bar m_1}+1)}{2}+m_2{\bar m}_2)C_4C_4\nonumber
\\ && -(\frac{m_2(m_2+1)}{2}+\frac{{\bar m}_2
({\bar m}_2+1)}{2}+m_1{\bar m}_1)S_4S_4\nonumber
\\&& +(n{\bar m}_1+{\bar n}m_1)Q_s+
(n{\bar m}_2+{\bar 
n}m_2)Q_s'\,\,\,\,\,\,\,\,\,\,\,\,\,\,\,\,\,\,\,\,\,\,\,\,\,\,\,\,\,\,\,
\eea
Hence, the gauge group is $U(16)_9\otimes U(8)_5\otimes
U(8)_{\overline{5}}$. The $U(16)$ arises on D9 branes while the $U(8)_5$
and $U(8)_{\overline{5}}$ live on D5-branes and ${\bD 5}$-branes located
at $y= \pi R$  and  $y=0$, respectively. The massless spectrum is given
in Table 3.

\begin{table}[h]
\begin{center}
\renewcommand{\arraystretch}{0.3}
\begin{tabular}{  | c || l |}
  \hline
&  \\ & Representation of $  U(8)_{\overline{5}}
\otimes U(16)_9 \otimes U(8)_5$ \\ &  \\
  \hline \hline
&  \\
gauge  bosons:&$({\bf 64},1,1)$ + $(1,{\bf 256},1)$
+ $(1,1,{\bf 64} )$ \\ &  \\ \hline &  \\
fermions: &$ ({\bf 64},1,1)_- + (1,1,{\bf 64} )_+$ \\ &  \\  &
$({\bf 36},{\bf 1},{\bf 1})_+ + ({\bf\overline{ 36}},{\bf 1},{\bf 1})_+$ +
$({\bf 1},{\bf 1},{\bf 36})_-  +
({\bf 1},{\bf 1},{\bf \overline{36}})_-$
\\ &  \\ & $({\bf{\bar 8}}, {\bf 16},{\bf 1})_- +
({\bf 1},{\bf \overline{16}},{\bf 8})_+$
\\ & \\ \hline &  \\
scalars:& $ 4 \times [({\bf 1},{\bf 120},{\bf 1})+
({\bf 1},{\bf \overline{120}},{\bf 1})]$
\\  &  \\ &
$ 4\, \times \, [({\bf 28},{\bf 1},{\bf 1})+
({\bf \overline{28}},{\bf 1},{\bf 1}) + ({\bf 1},{\bf 1}, {\bf 28}) +
({\bf 1},{\bf 1}, {\bf \overline{28}})]$ \\ &  \\ &
  $ 2 \times [ ({\bf{\bar 8}},{\bf 16},{\bf 1}) +
({\bf 1},{\bf \overline{16}},{\bf 8})]$
\\ &  \\ \hline
\end{tabular}
\end{center}
\caption{Massless spectrum in six dimensions. The indices $+$ and $-$ 
of fermions represent the six-dimensional chiralities. 
}\end{table}

Let us enumerate the main features of the spectrum.
First, we discuss the quantum numbers of the D9 brane fermions.
These are not modified by the interchange of branes with anti-branes.
As discussed in section 3, the Scherk-Schwarz boundary conditions keep
only those with one chirality at $y=0$ and those with the opposite
chirality at $y= \pi R$. Our convention is to denote the chirality of
(even) gauginos at $y=0$ by $+$ and the one of (odd) gauginos at $y=\pi
R$ by $-$, as done in Tables 2 and 3. The supersymmetry  transformations
allow to derive the chiralities of all other bulk fermions, that can be
also read off from the partition function. They are listed in Table 2. As
a consequence  of the fact that hypermultiplets  and gauginos should have
opposite chirality, the ND fermions have chirality $-$ at $y=0$ and $+$
at $y=\pi R$. Note also  that these ND states appear as half of
hypermultiplets because of the pseudo-reality condition.

Second, we discuss the quantum numbers of massless states living on the
D5 and $\bD 5$ branes. The chirality flip of the fermions when the
position of the branes and anti-branes are interchanged is
described in Table 2. There is no more (linear) supersymmetric couplings
between the bulk states (as the gravitinos) and those on the D5 and
${\bD 5}$ branes. A more striking signal of supersymmetry breaking
is that matter  fermions and scalars do not form anymore hypermultiplets.
While the  scalars  still transform  in the anti-symmetric representation
of the gauge group, the fermions transform now in the symmetric
representation. This is a consequence of a sign change in the M{\"o}bius
amplitude due to the new orientifold projection.

Third, the supersymmetries $Q_e$ and $Q_o$ are non-linearly realized
on the $\bD 5$ and D5 branes located at $y=0$ and $y=\pi R$, respectively.
We focus here on $y=0$ boundary, the situation at $y=\pi R$ being similar
and can be obtained by a chirality flip. Note first that at $y=0$, $Q_o$
is broken  by the orbifold and does not play any role. Note also that
although there are fermions in the adjoint representation of $U(8)$,
these do not have the proper chirality to be  the gauginos superpartners
of the gauge bosons of $U(8)$ under $Q_e$. In fact, their representation
is reducible, as $64=63+1$  which corresponds to the gauge group
decomposition $U(8) \rightarrow  SU(8) \otimes U(1)$. The singlet fermion
can be identified with the  Goldstino of the non-linearly realized
supersymmetry $Q_e$ on the $\bD 5$ world-volume, from the point of view of the 55 matter fermions living on the 5-branes. Its associated
Chan-Paton  factor is given by $\frac{1}{\sqrt{8V_5}} {\bf 1}_8$ with ${\bf
1}_8$  representing the identity matrix of rank 8. The decay constant of
the Goldstino can be computed from the scattering of two Goldstinos with
two matter fermions on the $\bD 5$ branes and equals $8 \tau_5$ with
$\tau_5=\frac{1}{16\pi^4}\frac{M_s^6}{g_s}$ the $\bD 5$ brane tension. When the matter fermions are 95 strings (localized on the brane intersection), the Goldstino appears as a linear combination of 99 and 55 states, as we described in section 2.

Finally, as a consistency check, one can compute the anomaly polynomial:
\bea
A=\frac{1}{4}(trF_5^2-trF_{\overline{5}}^2)(trF_9^2-\frac{1}{2}trR^2)
\eea
whose factorized form allows for a generalized Green-Schwarz anomaly 
cancellation mechanism \cite{GS,sa6,SW}, 
where the gauge anomaly is cancelled by adding a Wess-Zumino counterterm
in the Lagrangian \cite{sa6}.

\section{Chiral four-dimensional example }

Starting from the  six-dimensional model of section 5, it is possible 
to construct a  chiral four-dimensional descendant. As an example, a 
compactification on $T^2$ followed by an additional orbifolding by a
${\mathbb {Z}_3 }
$  discrete symmetry leads to a chiral type IIB orientifold on 
$T^6 / ({\mathbb {Z}_2 }
\times {\mathbb {Z}_3 }) \equiv T^6 / {\mathbb {Z}^{\prime}_6 }$ \cite{ads4} 
whose  the Yang-Mills and  matter field content will be derived below.
 
Consider the ten-dimensional type IIB string theory  
compactified on $ T^6 / {\mathbb {Z}^{\prime}_6 }$. The 
internal  $T^6 $ dimensions $ \{ x^4, \cdots,x^9\equiv y \} $ 
are paired  into three complex variables, $z_1= x^4 + i x^5$, 
$ z_2= x^6 + i x^7$ and   $z_3= x^8 + i y$. The  actions of the
discrete groups $\mathbb { Z}_2$ and $\mathbb { Z}_3$ on the space-time
coordinates are defined as :
\bea
{\mathbb Z}_2 \, : && z_1 \rightarrow z_1, \, \, \, \, \qquad  z_2 
\rightarrow - z_2, \qquad  z_3 \rightarrow - z_3 \\
{\mathbb Z}_3 \, : &&z_1 \rightarrow \omega^{-1} z_1, \qquad  z_2 
\rightarrow   \omega z_2, \qquad  z_3 \rightarrow  z_3  
\label{action}
\eea
where  $\omega=e^{\frac{2i\pi}{3}}$. The $\mathbb { Z}_2$ action leads to 
the six-dimensional  model of section 5.
Since the additional $\mathbb {Z}_3$ projection does not introduce any new
D-brane sector, the four-dimensional open string spectrum can
be obtained by an appropriate truncation of the spectrum of the
six-dimensional $T^4 /{\mathbb {Z}_2 }$ model, we constructed in the
previous section, upon a further toroidal compactification on $T^2$ to
four dimensions. Besides its space-time action (\ref{action}), 
$\mathbb{Z}_3$ acts also on the Chan-Paton factors in a way dictated by
tadpole cancellation conditions \cite{zsix,ads4}.

On the one hand, the model contains D9 branes that extend in the whole 
bulk. The associated DD string fermionic modes feel the Scherk-Schwarz 
breaking and acquire mass shifts  by $1/2R$. On the other hand, it
contains ${\bD 5}$ and D5 branes located at $y=0$ and $y= \pi R$
respectively, with world-volumes transverse to the $y$ direction. As a
consequence, the massless modes  arising from open strings with at least
one end on the D5 or ${\bD 5}$ branes are unaffected at tree-level.
Performing a $T$-duality along the coordinate $z_1$, this system is
transformed into a  configuration of D7-branes extending in $z_2$ and
$z_3=x^8+iy$, together with ${\bD 3}$ and D3 branes located at $y=0$ and
$y= \pi R$, respectively. The Scherk-Schwarz breaking acts now, besides
the closed string bulk, on the fermions  propagating on the 
world-volume of the D7-branes.

From the $\mathbb {Z}_3$ action on the Chan-Paton factors, we can derive
its action on the fundamental representations of $U(16)$ and $U(8)$ and
furthermore on all states. Indeed, all other representations  can be
obtained from tensor products of fundamentals with  anti-fundamentals.
Labelling the representations with a subscript $\omega^k,\,  k=0,\pm 1 $
that indicates how they transform  under $\mathbb {Z}_3$, we have: 
\bea
{\bf 16}&=&{\bf (4,1,1)_{\omega}+(1,4,1)_{\omega^{-1}}+(1,1,8)_{1}}\nonumber \\
{\bf 8}&=&{\bf (2,1,1)_{\omega}+(1,2,1)_{\omega^{-1}}+(1,1,4)_{1}}
\eea
The tensor product $8 \otimes \overline{8}$ leads then to the following
decomposition of the adjoint and 2-index symmetric and 
antisymmetric representations:
\bea
{\bf 64}\!\!\!&=&\!\!\!\!{\bf (4,1,1)_1+(1,4,1)_1+(1,1,16)_1+(2,\overline{2},1)_{\omega^{-1}}+(\overline{2},2,1)_{\omega}+(2,1,\overline{4})_{\omega}} \nonumber \\ && 
\qquad \qquad \qquad \qquad \qquad \qquad \qquad\!\!\!\!{\bf  +(\overline{2},1,4)_{\omega^{-1}}+(1,2,\overline{4})_{\omega^{-1}}+(1,\overline{2},4)_{\omega}} 
 \nonumber \\ 
{\bf 28}\!\!\!&=&\! \!\!\!{\bf (1,1,1)_{\omega^{-1}}\!+(1,1,1)_{\omega}+(1,1,6)_1+(2,2,1)_1\!+
(2,1,4)_{\omega}+(1,2,4)_{\omega^{-1}} } \nonumber \\
{\bf 36}\!\!\!&=&\!\!\!\!{\bf (3,1,1)_{\omega^{-1}}\!+(1,3,1)_{\omega}\!+(1,1,10)_1\!+(2,2,1)_1+\!(2,1,4)_{\omega}\!+(1,2,4)_{\omega^{-1}}} \nonumber 
\eea
Combining these transformations with the $\mathbb {Z}_3$ action on the 
space-time quantum numbers (\ref{action}), we can identify the massless
spectrum as the set  of states left invariant by the product of the two
actions. We find that, the $\mathbb {Z}_3$ action breaks each of the 
six-dimensional gauge group factors into three subgroups:
\bea
{ {\bD 5}}:&& U(8)_{\overline{5}} \rightarrow U(2)_{\overline{5}}
 \otimes U(2)_{\overline{5}} \otimes U(4)_{\overline{5}} \nonumber \\
{\rm D9}:&& U(16)_9 \rightarrow U(4)_9 \otimes U(4)_9 \otimes U(8)_9  \nonumber \\
{\rm D5}:&& U(8)_5 \rightarrow U(2)_5 \otimes U(2)_5 \otimes U(4)_5 
\eea 

The full open string massless spectrum is listed in Table 4. Note that
each 5-brane sector contains a Pati-Salam type gauge group with two
chiral families with the quantum numbers of quarks and leptons and two
electroweak Higgs bosons. A three family model can be obtained by
breaking for instance the three $U(4)$ factors 
$U(4)_{\overline{5}}\otimes  U(4)_9\otimes U(4)_9$ into the diagonal
$U(4)$ subgroup, using the scalar bi-fundamental representations
available in the massless spectrum. As a result, one obtains an
additional chiral family from the ${\overline{5}}9$ sector.

\begin{table}[pht!]
\begin{center}
\renewcommand{\arraystretch}{0.3}
\begin{tabular}{  | c || l |} 
 \hline
& \\${\overline{5}}{\overline{5}}$  & Representation of $  U(2)_{\overline{5}}
\otimes U(2)_{\overline{5}} \otimes U(4)_{\overline{5}}$ \\ &  \\
 \hline \hline
&  \\
gauge  bosons:& ${\bf (4,1,1)+(1,4,1)+(1,1,16)}$
 \\ &  \\ \hline &  \\
fermions: &${\bf(4,1,1)+(1,4,1)+(1,1,16)}$ \\ &  \\ &
$ {\bf (3,1,1)+  (1,\overline{3},1)+(1,1,10)+(1,1,\overline{10})}$ \\ &  \\ &
${\bf (2,1,\overline{4})+ (\overline{2},1,\overline{4})+
(1,2,4)+ (1,\overline{2},4)}$ \\ &  \\ &
${\bf (2,2,1)+ (\overline{2},2,1)+ (\overline{2},\overline{2},1) }$ 
\\ & \\ \hline &  \\
scalars:& ${\bf (1,1,1) +  (1,\overline{1},1) + (1,1,6) + 
(1,1,\overline{6})}$   
\\  &  \\ &
${\bf (2,2,1) +  (\overline{2},\overline{2},1) +(1,2,4)+  
(\overline{2},1,\overline{4})}$
\\ &  \\ \hline \mco{2}{c}{}\\  \mco{2}{c}{}\\  \mco{2}{c}{}\\  \hline
&  \\ ${\overline{5}}9$ & Representations of \\ &  \\ 
& $ U(2)_{\overline{5}}
\otimes U(2)_{\overline{5}} \otimes U(4)_{\overline{5}}
 \otimes  U(4)_9 
\otimes U(4)_9\otimes U(8)_9 $ \\ &  \\
 \hline  \hline
 &  \\
``Chiral multiplets'': & ${\bf (1,2,1;1,4,1) + 
(1,1,4;4,1,1) + (2,1,1;1,1,8)}$ \\ &  \\ 
& ${\bf (\overline{2},1,1;\overline{4},1,1)+ 
(1,1,\overline{4};1,\overline{4},1)
+ (1,\overline{2},1;1,1,\overline{8})} $
\\ &  \\ \hline \mco{2}{c}{}\\  \mco{2}{c}{}\\  \mco{2}{c}{} \\ \hline
& \\$99$  & Representation of $  U(4)_9
\otimes U(4)_9 \otimes U(8)_9$ \\ &  \\
 \hline \hline
&  \\
gauge  bosons:& ${\bf (16,1,1)+(1,16,1)+(1,1,64)}$
 \\ &  \\ \hline &  \\
scalars:&$ {\bf (6,1,1)+  (1,\overline{6},1)+(1,1,28)+(1,1,\overline{28})}$ \\ &  \\ &
${\bf (4,1,\overline{8})+ (\overline{4},1,\overline{8})+
(1,4,8)+ (1,\overline{4},8)}$ \\ &  \\ &
${\bf (4,4,1)+ (\overline{4},4,1)+ (\overline{4},\overline{4},1) }$ 
\\ & \\ \hline
\end{tabular}
\end{center}
\caption{Massless spectrum in four dimensions. All the scalars are complex 
and the fermions left-handed. The representations with a bar on top
have opposite $U(1)$ charges to those without bar. The 55 and 59
states are deduced by simple conjugation from the
${\overline{5}}{\overline{5}}$ and  
${\overline{5}}9$ representations, respectively. }\end{table}

On the other hand, the massless closed string spectrum contains besides
the graviton, dilaton and axion, 5 untwisted complex scalars, 18 
$\mathbb{Z}_3$ twisted complex scalars and 24 twisted chiral multiplets
localized in the $y$-direction.

We discuss now the main aspects of supersymmetry breaking
in these models.
As we explained above, the effect of $\mathbb {Z}_3$ orbifolding amounts
to an additional projection that breaks half of the non-linear (or
linear) supersymmetry of the six-dimensional model, in the presence (or
absence) of the Scherk-Schwarz deformation.
The non-linear realization of supersymmetry on the boundaries 
is therefore inheritated from the six-dimensional model. Let us denote by 
$Q_e^{(6)}$ and $Q_o^{(6)}$ the two supersymmetric generators of the 
six dimensional supersymmetries non-linearly realized at $y=0$ and 
$y=\pi R$, respectively. Each of them has 8 spinorial 
degrees of freedom. The $\mathbb {Z}_3$ action projects away half
and we are left with two four-dimensional supercharges, $Q_e^{(4)}$ and
$Q_o^{(4)}$ at $y=0$ and $y=\pi R$, respectively. This is described as:
\bea
Q_e^{(6)}= Q_{e,1}^{(4)}+ Q_{e,2}^{(4)}
&\stackrel{\mathbb {Z}_3}{\longrightarrow}&Q_{e,1}^{(4)}\equiv Q_e^{(4)} 
\label{decQ}\\
Q_o^{(6)}= Q_{o,3}^{(4)}+ Q_{o,4}^{(4)}
&\stackrel{\mathbb {Z}_3}{\longrightarrow}&Q_{o,3}^{(4)}\equiv Q_o^{(4)} 
\eea
It is important to stress that the non-linearly realized supersymmetries 
on the world-volumes of the ${\bD 5}$  and the D5 branes located at 
$y=0$ and  $y= \pi R$, respectively, are different:   
$Q_e^{(4)}\neq Q_o^{(4)}$. Also, as in the six-dimensional case, the
fermions  in the adjoint representations can not be identified with
gauginos as  they hold wrong chiralities.

It is now straightforward to identify the Goldstino field associated
with  the non-linear realization of these supersymmetries on the
boundaries. For simplicity, we will restrict our discussion to the
boundary $y=0$. From eq. (\ref{decQ}), we see that  the Goldstino of
$Q_e^{(4)}$ is a component of the  six-dimensional Goldstino of
$Q_e^{(6)}$.

First, we remind that in six dimensions the Goldstino is identified
as the singlet component in the decomposition $64=63+1$ of the adjoint 
representation of $U(8)= SU(8) \otimes U(1)$. We denote this singlet as
$\lambda_-^{(6)}$. Under ${\mathbb Z}_3$, this is decomposed as:
\bea
\lambda_{-}^{(6)}=\lambda_{-,P}^{(4)}+\lambda_{+,I}^{(4)}
\eea
where the component $\lambda_{-,I}^{(4)}$ is projected out
while $\lambda_{-,P}^{(4)}\equiv \chi$ remains and it is identified with
the Goldstino in four dimensions.

Second, note that the overall $U(1)$ factor in six dimensions is
decomposed into three $U(1)$'s by ${\mathbb Z}_3$, as:
\bea
{\bf 64} \longrightarrow {\bf (3+1,1,1)+(1,3+1,1)+(1,1,15+1)}
\eea
where we identify three  singlets $\lambda_{\ov{5},i}^{(4)}, \, i=1,2,3$ 
\bea
{\bf 4=3+(1} \leftarrow \lambda_{\ov{5},1}^{(4)},\lambda_{\ov{5},2}^{(4)}) \\
{\bf 16=15+(1} \leftarrow \lambda_{\ov{5},3}^{(4)})
\eea
One linear combination of the  singlets corresponds to the Goldstino
field $\chi=\lambda_{-,P}^{(4)}$. Taking into account the normalization
of the Chan-Paton generators as in section 2, the Goldstino is given by\footnote{Moreover, one should take the linear combination of 99 and 55 states, as discussed in section 2.}:
\bea
\chi = \frac{1}{2} [ \lambda_{1}^{(4)}+ \lambda_{2}^{(4)}+ \sqrt{2}
\lambda_{3}^{(4)} ]\, .
\eea

Finally, let us discuss the issue of $U(1)$ anomalies in this model.
There are nine $U(1)$ factors 
whose associated charges are denoted in a self-explanatory notation by
$ Q_i^\alpha$, where $\alpha ={\overline{5}}, 9, 5 $ labels the different
kind of  branes and $i=1,\cdots,3$ counts the $U(1)$'s for given
$\alpha$. These $U(1)$'s have mixed anomalies from triangular diagrams
with two  non-abelian gauge bosons from $SU(2)$ or $SU(4)$ factors.
Following ref. \cite{uran}, we denote by $A_{ij}^{\alpha\beta}$ the
associated anomaly coefficient $Tr (Q_i^\alpha T_j^\beta T_j^\beta)$,
with $T_j^\beta$ the generator of  the non-abelian group $j$ from branes
of the type $\beta$. These anomalies can be collected in a matrix:  
\begin{displaymath}
\mathbb A=(A_{ij}^{\alpha\beta})_{\stackrel{\alpha,\beta\,\,=\,{\bar 5},9,5 }{1\leq\,\, i,j \,\,\leq 3}}=\left( \begin{array}{ccccccccc} 
\, \,  1 &\,\,   1 &\! -4 &\! -2 &\, \,  0 &\, \,  4 &\, \,  0 &\, \,  0 &
 \, \,  0 \\ 
\! -1 &\! -1 &\, \,  4 &\, \,  0 &\, \,  2 &\! -4 &\, \,  0 & \, \, 0 &\, \,  0 \\
\, \,  0 &\, \,  0 &\, \,  0 &\, \,  2 &\! -2 &\, \,  0 & \, \, 0 &\, \,  0 
&\, \,  0 \\
\! -1 &\, \,  0 &\, \,  2 &\, \,  0 &\, \,  0 &\, \,  0 &\, \,  1 &\, \,  0 &\! 
-2 \\
\, \,  0 &\, \,  1 &\! -2 &\,  \, 0 &\,  \, 0 &\,  \, 0 &\,  \, 0 &\! -1 &\,  
\, 2 \\
\,  \, 1 &\! -1 &\, \,  0 &\, \,  0 &\, \,  0 &\, \,  0 &\! -1 &\, \,  1 &\, \, 
 0 \\
\, \,  0 &\, \,  0 &\, \,  0 &\, \,  2 &\, \,  0 &\! -4 &\! -1 &\! -1 &\, \,  4 \\
\, \,  0 &\, \,  0 &\,  \, 0 &\, \,  0 &\! -2 &\, \,  4 &\, \,  1 &\, \,  1 &\! 
-4 \\
\, \,  0 &\, \,  0 &\, \,  0 &\! -2 &\, \,  2 &\, \,  0 &\, \,  0 &\,  \, 0 &\, \, 
 0 \\
\end{array}
\right)
\end{displaymath}
From the matrix anomaly $\mathbb A$, we can deduce the anomaly free
$U(1)$'s which are given in an appropriate basis by the linear
independent combinations:
\begin{displaymath}
\begin{array}{c} 
Q_1^{\overline{5}}+Q_2^{\overline{5}}+\frac{1}{2}Q_3^{\overline{5}}\\ 
Q_1^9+Q_2^9+\frac{1}{2}Q_3^9 \\
Q_1^5+Q_2^5+\frac{1}{2}Q_3^5 \\ 
Q_1^{\overline{5}}-Q_2^{\overline{5}}+Q_1^5-Q_2^5 \\
Q_1^{\overline{5}}+Q_2^{\overline{5}}-4Q_3^{\overline{5}}+Q_1^9+Q_2^9-4Q_3^9+Q_1^5+Q_2^5-4Q_3^5  \\
\end{array}
\end{displaymath}
The remaining $U(1)$'s are anomalous and can be expressed in a convenient
basis by:
\begin{displaymath}
\begin{array}{c} 
Q_1^{\overline{5}}-Q_2^{\overline{5}}-Q_1^5+Q_2^5 \\
Q_1^9-Q_2^9 \\ 
Q_1^{\overline{5}}+Q_2^{\overline{5}}-4Q_3^{\overline{5}}
-Q_1^5-Q_2^5+4Q_3^5  \\
Q_1^{\overline{5}}+Q_2^{\overline{5}}-4Q_3^{\overline{5}}-2Q_1^9-2Q_2^9+8Q_3^9+Q_1^5+Q_2^5-4Q_3^5  \\
\end{array}
\end{displaymath}
All these anomalies are expected to be cancelled by a generalized
Green-Schwarz mechanism, which introduces 4 axions transforming
non-trivially under the corresponding anomalous gauge symmetries.

\section{Conclusion}

A simple and elegant way to break supersymmetry on D-branes is to place
them on top of orientifold planes that preserve different
supersymmetries. The resulting spectra are not superymmetric but
supersymmetry is still realized non-linearly with a Goldstino living on
the D-brane world-volume. A particular advantage of this method is the
absence of tachyons. However, previous implementations of this idea 
suffered from a non-vanishing tree-level cosmological constant of the
order of the (string scale)$^4$, which makes any quantitative prediction
and computation questionable.

In this work, we propose a solution to this problem. Our construction
relies on the possibility to use at the same time a Scherk-Schwarz
deformation, which introduces an additional ``tiny" breaking of
supersymmetry in the bulk, that vanishes in the decompactification limit.
An immediate consequence of the Scherk-Schwarz boundary conditions is the
introduction of an additional set of anti-branes with orientifolds,
so that the total contribution to the vacuum energy vanishes.

In this paper, we have computed the four-fermion effective interaction of
the Goldstino with matter fermions. In the simplest case, where matter fermions correspond to open strings with both ends on the same set of branes, we found that the Goldstino decay constant is given
by the total tension of the corresponding D-branes. The 4-fermion interactions involve also a dimensionless parameter
that takes two different values, depending on whether the matter fermions
 have the same or different internal helicity with the Goldstino.

As we have shown in explicit examples, our construction allows to
obtain non-supersymmetric models with  chiral spectrum and interesting
gauge groups, such as Pati-Salam type with three generations of
quarks and leptons. These constructions are very useful in the context of
low-scale string theory with a string tension lying in the TeV region. On
the other hand, more general constructions allow for the simultaneous
presence of branes with non-supersymmetric world-volume and others with
(tree-level)  supersymmetric massless modes. In these models, there is
another option that the Standard Model may reside on one of the
supersymmetric branes, in which case the string scale should be much
larger than the TeV, for instance at intermediate energy scales.

Our construction also provides a consistent framework for investigating
properties of non-supersymmetric brane-worlds, such as threshold
corrections to gauge couplings, and mediation of supersymmetry breaking.
We plan to return to these issues in the near future.

\section*{Acknowledgements}

This work was supported in part by the European Commission under RTN
contract HPRN-CT-2000-00148, and in part by the INTAS contract 99-1-590. 
The work of K.B. is supported by the EU fourth framework program 
TMR contract FMRX-CT98-0194 (DG 12-MIHT). 
A. L. thanks the Theory Division of CERN for its hospitality and partial 
financial support.


\appendix 

\section{Notations for one-loop amplitudes}
For $10-d$ compact space dimensions, the one-loop string amplitudes 
${\bf T}$, ${\bf K}$, ${\bf A} $, $ {\bf M}$ corresponding to the 
torus, Klein bottle,   annulus and  M{\"o}bius  diagrams, respectively, 
are given by: 
\bea
{\bf T}&=&{1 \over ({4 \pi^2 {\alpha'}})^{d \over 2}}  
\int \frac {d \tau d {\bar \tau}}{  ({\rm Im \ \tau})^{1+{d \over 2}}} 
 \, \, 
| \frac {1}{\eta(\tau)}|^{2d-4}\, \, 
{\cal T} \, ,
\\
{\bf K} &=& {1 \over  ({4 \pi^2 {\alpha'}})^{d \over 2}} 
\ \int_0^{\infty} {d \tau_2 \over \tau_2^{1+{d \over 2}}} 
\frac {1  }{\left[\eta(2 i \tau_2)\right]^{d-2}}\, \,  {\cal K}
\nonumber \\  &=&
{1 \over  ({4 \pi^2 {\alpha'}})^{d \over 2}} \int_0^{\infty}
dl \ \frac {1  }{\eta(i l)^{d-2}}
\, \, {\tilde{\cal K}}\ , \nonumber 
\eea
\bea
{\bf A}& =& {1 \over  ({8 \pi^2 {\alpha'}})^{d \over 2}} 
\int_0^{\infty} {d t \over t^{1+{d \over 2}}} 
 \frac {1  }{\left[\eta(\frac{i t}{2})\right]^{d-2}}\, \,  {\cal A} 
\nonumber \\  &=& {1 \over  ({8 \pi^2 {\alpha'}})^{d \over 2}}
\int_0^{\infty} dl \ \frac {1  }{\eta(i l)^{d-2}}\, \,  {\tilde{\cal A}}\ , 
\nonumber \\
{\bf M} &=&   {1 \over  ({8 \pi^2 {\alpha'}})^{d \over 2}}
\int_0^{\infty} {d t \over t^{1+{d \over 2}}}
\frac {1  }{\left[\hat{\eta}(\frac{i t}{2}+\frac{1}{2})\right]^{d-2}}
\, \,  {\cal M}
\nonumber \\  &=&  {1 \over  ({8 \pi^2 {\alpha'}})^{d \over 2}}
 \int_0^{\infty} dl \ \frac {1  }{\hat{\eta}(i l+\frac{1}{2})^{d-2}}\, \,  {\tilde{\cal M}}
\nonumber 
\eea 
where $\alpha' = M_s^{-2}$.  The integral over the modular parameter 
$\tau = \tau_1 + i\tau_2$ in the torus amplitude is performed over the fundamental domain:
\bea
 -\frac {1}{2} \leq \tau_1 \leq \frac {1}{2}
\ , \, \tau_2 \geq 0 \ , \ |\tau| \geq 1 \ 
\eea
The direct (open string) channel amplitudes $ {\cal T}$, $ {\cal K}$, $ {\cal A}$ and 
$ {\cal M}$ are given as functions of the conformal field theory characters 
 and the compactification lattice sums  defined below. The 
corresponding expressions in the transverse (closed string) channel $ {\tilde {\cal K}}$, 
${\tilde {\cal A}}$ and ${\tilde {\cal M}}$ are obtained by the 
following transformations on the integration variables
\bea 
{\bf K} &:& \qquad\qquad \qquad 2 \tau_2  \ \, \,
\stackrel{ S} {\longrightarrow}  \qquad
 \frac {1} { 2 \tau_2}\equiv l \\
{\bf A} &:& \qquad\qquad \qquad \frac {t}{2} \ \, \,
\stackrel{ S} {\longrightarrow}  \qquad
\frac{2}{t} \equiv l \\
{\bf M} &:& \qquad\qquad {it \over 2}+{1 \over 2} \ 
\stackrel{P}{\longrightarrow}   \qquad
{i \over 2t}+{1 \over 2} \equiv il + {1 \over 2} \ . \label{dirtotrans}
\eea
and allow to express  one-loop open string 
diagrams  as tree-level  closed string ones.
The conformal field theory characters are given by 
\bea
\chi_r = q^{h_r -{c \over 24}} \sum_{n=0}^{\infty} d_n^r q^n \ , 
\eea
where $h_r$ is the conformal weight, $c$ is the central
charge of the conformal field theory and the $d_n^r$ are positive
integers. The hatted characters  are defined as:
\bea 
{\hat\chi}_r \ (il+\frac{1}{2})=e^{-i\pi h_r} \chi_r \ (il+\frac{1}{2})\ .
\label{hatdef}
\eea 
The above characters can be expressed using the  Dedekind $\eta$ function  
\bea
\eta(\tau) = q^{1\over 24} \prod_{n=1}^\infty (1-q^n)\ , \label{a1}
\eea 
and the Jacobi $\theta$ theta functions with general characteristic 
$(\alpha,\b)$: 
\bea
\vartheta \left[{\alpha }\atop{ \b }\right] (z,\tau) = \sum_{n\in Z}
e^{i\pi\tau(n-\alpha)^2} e^{2\pi i(z- \b)(n-\alpha)} \ ,
\eea
where   $q=e^{2\pi i\tau}$.
We use the notation:
\bea
\vartheta_1(z,\tau) &\equiv&  \vartheta \left[{{1\over 2} \atop {1\over
2} } \right] (z,\tau),\quad 
\vartheta_2(z,\tau) \equiv  \vartheta \left[{{1\over 2} \atop 0 }\right]
(z,\tau), \\ 
\vartheta_3(z,\tau) &\equiv&  \vartheta \left[{0 \atop 0
    }\right](z,\tau), 
\quad 
\vartheta_4(z,\tau) \equiv  \vartheta \left[{0 \atop {1\over 2} }\right]
(z,\tau)
\eea
In the orbifold models we consider in this work, the partition function can be expressed in terms of the  $SO(2n)$ characters 
\bea 
I_{2n} &=& {1 \over 2 \eta^n} ( \theta_3^n + \theta_4^n) \ , \qquad\quad 
V_{2n}={1 \over 2 \eta^n} (
\theta_3^n - \theta_4^n) \ , \nonumber \\ S_{2n} &=& {1 \over 2 \eta^n} ( \theta_2^n +
i^n
\theta_1^n) \ , \qquad C_{2n}={1 \over 2 \eta^n} 
( \theta_2^n - i^n \theta_1^n) 
\label{E1}
\eea 
At the lowest level, 
$I_{2n}$ represents a
scalar, $V_{2n}$  a vector, while  $S_{2n}$, $C_{2n}$
represent spinors of opposite chiralities. The transformations 
$S$ and $P$ defined in eq. (\ref{dirtotrans}) from direct to transverse
channels act on the vector $\{I_{2n},V_{2n},S_{2n},C_{2n}\}$ 
through the matrices:

\bea 
S_{(2n)} =  {1 \over 2} \left(
\begin{array}{cccc}  1 & 1 & 1 & 1 
\\ 1 & 1 & -1 & -1 
\\ 1 & -1 & i^{-n} & -i^{-n} 
\\ 1 & -1 & -i^{-n} & i^{-n}
\end{array}
\right) \ , \  P_{(2n)} = \left(
\begin{array}{cccc}  c & s & 0 & 0 \\  
s & -c & 0 & 0 \\ 
0 & 0 & \zeta c & i \zeta s \\
0 & 0 & i \zeta s & \zeta c
\end{array}\right) 
\eea   
where $c= \cos ({n \pi /4})$, $s= \sin ({n \pi /4})$ and 
$\zeta= e^{-i{n \pi/4}}$. 

Useful combinations of these characters in the case of
compactifications on $T^4/{\mathbb Z}_2$ are:

\bea 
Q_o = V_4I_4-C_4C_4 \ , \qquad Q_v = I_4V_4-S_4S_4 \ , \nonumber\\ 
Q_s = I_4C_4-S_4I_4 \ , \qquad Q_c = V_4S_4-C_4V_4 \ ,
\eea 
and
\bea 
Q'_o = V_4I_4-S_4S_4 \ , \qquad Q'_v = I_4V_4-C_4C_4 \ , \nonumber\\ 
Q'_s = I_4S_4-C_4I_4 \ , \qquad Q'_c = V_4C_4-S_4V_4 \ ,  
\eea 
Here the first factor refers to the six-dimensional 
 space-time  (in the light-cone gauge), while the second refers to 
the internal compact space. 

\section{Compactification lattice summations}
The contributions to the one-loop amplitudes from the compactification lattice
can be expressed as function of the series $ Z_{m+a,n+b}$. For instance, the torus amplitude ${\cal T}$ can be expressed as function of: 
\bea 
Z_{m+a,n+b}(\tau,{\bar \tau})= {1 \over
|\eta(\tau)|^2} q^{({{m+a} \over R}+{(n+b)R \over 2})^2} {\bar
q}^{({{m+a} \over R}-{(n+b)R \over 2})^2} 
\eea 
The other  one-loop diagrams are expressed as  functions of
Kaluza-Klein and winding lattice summations   
\bea 
Z_{m+a}(\tau) = \frac{q^{{1\over 2} {\left(\frac{m+a}{R}\right)}^2}}{\eta(\tau)}\ ,
\qquad
\tilde{Z}_{n+ b}(\tau) =\frac{q^{{1\over 2} {\left( (n + b){R\over 2}
\right)}^2}}{\eta(\tau)} \ . 
\eea 
Under Poisson resummation we have
\bea
\sum_m e^{2i\pi m b} Z_{m+a} (-{1 \over \tau})=R\ e^{-2i\pi ab} \sum_n e^{-2i
\pi na} {\tilde Z}_{2n+2b} ({\tau})  
\eea 
\bea
\sum_{m,n} Z_{m,n}(\tau,{\bar \tau})&=& {R{\tau_2}^{-1/2} \over{\sqrt 2} |\eta(\tau)|^2} \sum_{{\tilde m},n} e^{-{\pi
R^2 \over 2 \tau_2} |{\tilde m}+n \tau|^2 }
\eea
It is sometimes convenient to use  the projected lattice sums 
\bea 
E_0 &=& \sum_{m,n} {1 + (-1)^m \over 2} Z_{m,n} \ ,\qquad\qquad
O_0 =
\sum_{m,n} {1 - (-1)^m \over 2} Z_{m,n} \ , \nonumber \\ E_{1/2} &=& \sum_{m,n}
{1 + (-1)^m
\over 2} Z_{m,n+1/2} \ ,\qquad O_{1/2} = \sum_{m,n} {1 - (-1)^m \over 2}
Z_{m,n+1/2} \quad ,\nonumber 
\eea
and 
\bea 
E^{\prime}_0 &=& \sum_{m,n} {1 + (-1)^n \over 2} Z_{m,n} \ ,\qquad\qquad
O^{\prime}_0 =
\sum_{m,n} {1 - (-1)^n \over 2} Z_{m,n} \ , \nonumber \\ 
E^{\prime}_{1/2} &=& \sum_{m,n}
{1 + (-1)^n
\over 2} Z_{m+1/2,n} \ ,\qquad O^{\prime}_{1/2} = \sum_{m,n} {1 - (-1)^n \over 2}
Z_{m+1/2,n} \quad ,\nonumber 
\eea 
where $E$ ($E^\prime$) and $O$ ($O^\prime$) refer correspondingly 
to even and odd 
KK momenta (windings) and the subscripts
$0$ and $1/2$ refer to unshifted and shifted winding (KK momenta). The 
primed sums are obtained from the unprimed ones through interchange of $m$ 
and $n$.
In the case of $D$ compact dimensions we will use the notation:
\bea
\Lambda^{(D)}&=& \prod_{i=1}^{D} \sum_{m_i}  Z_{m_i}\qquad 
{\tilde\Lambda}^{(D)}=\prod_{i=1}^{D} \sum_{n_i}  {\tilde Z}_{n_i} \nonumber \\
\Lambda^{(D,D)}&=& \prod_{i=1}^{D} \sum_{m_i,n_i}  Z_{m_i,n_i}\qquad 
\eea

\section{Supersymmetric one-loop partition functions in 9D and 6D}
For comparison, we provide the reader with the partition functions in the
supersymmetric cases in 9D and 6D.
With our conventions, the amplitudes for the supersymmetric
nine-dimen\-sional theory obtained by  compactification of the
$SO(32)$ type I strings on a circle are given by: 
\bea   {\cal T}&=&|V_8-S_8|^2
\sum_{m,n}Z_{m,n}\ , \qquad\qquad\qquad\ \    
{\cal K}={1 \over 2} (V_8-S_8) \sum_m Z_m
\ , \nonumber \\   
{\cal A}&=& {1 \over 2} (V_8-S_8) \sum_{i,j,m}
Z_{2(m+a_i+a_j)} \ , \qquad   {\cal M}=-{1 \over 2} ({\hat V}_8-{\hat
S}_8) \sum_{i,m} Z_{2(m+2a_i)}  \nonumber .  
\eea 
which
corresponds to two $O8$  orientifold planes with Ramond-Ramond (RR)
charge $-16$ and  32 $D8$-branes each carrying a unit of RR charge.

For the supersymmetric six-dimensional $T^4/{\mathbb Z_2}$ we have:
\bea 
{\cal T}&=&{1 \over 2} \Lambda^{(4,4)} |V_8-S_8|^2+ 
\frac{1}{2} |Q_o - Q_v|^2 {\biggl|\frac{2 \eta}{\theta_2}\biggr|}^4 
\nonumber +
\frac{1}{2} |Q_s + Q_c|^2 {\biggl|\frac{2 \eta}{\theta_4}\biggr|}^4 
\nonumber \\
&&+ \frac{1}{2} |Q_s - Q_c|^2 {\biggl|\frac{2 \eta}{\theta_3}\biggr|}^4 \ ,
\nonumber \\
{\cal K} &=& \frac{1}{4} \biggl\{ ( Q_o + Q_v ) ( \Lambda^{(4)} + 
{\tilde\Lambda}^{(4)}) + 
32 ( Q_s + Q_c ){\biggl(\frac{\eta}{\theta_4}\biggr)}^2 \biggr\}
\ , 
\nonumber 
\eea
\bea
{\cal A} &=& \frac{1}{4}  \biggl\{
(V_4 O_4 + O_4 V_4 - C_4 C_4 - S_4 S_4 ) 
\left[ n_{N}^2  \, \Lambda^{(4)} + n_{D}^2 \, {\tilde\Lambda}^{(4)} \right] 
 \nonumber \\ 
&& + (V_4 O_4 - O_4 V_4 - C_4 C_4 + S_4 S_4 ) 
\left( \frac{2 \eta}{\vartheta_2}\right)^2 
(R_{N}^2  + R_{D}^2 ) 
 \nonumber \\
&&\ + 2 (O_4 C_4 + V_4 S_4 - S_4 O_4 - C_4 V_4 ) 
\left( \frac{\eta}{\vartheta_4} \right)^2
 n_N n_D 
 \nonumber \\ 
&& + 2  (O_4 C_4 - V_4 S_4 - S_4 O_4 + C_4 V_4 ) 
\left( \frac{\eta}{\vartheta_3} \right)^2 
R_{N} R_{D} \biggr\} \ ,
\nonumber \\
{\cal M} &=& - \frac{1}{4} \biggl\{ (\hat V _4 \hat O_4 + \hat O_4
\hat V_4 - \hat C_4 \hat C_4 -\hat S_4 \hat S_4 ) 
\left[ n_N \, \Lambda^{(4)} +  n_D \,  {\tilde\Lambda}^{(4)} \right] 
\nonumber \\
&& - (\hat V _4 \hat O_4 - \hat O_4
\hat V_4 - \hat C_4 \hat C_4 + \hat S_4 \hat S_4 ) 
\left( \frac{2\hat\eta}{\hat \vartheta_2 } \right)^2
( n_N  + n_D ) 
\biggr\} \nonumber
\eea 
where the tadpole cancellation requires $n_N=n_D=32$ and $R_N=R_D=0$.

\section{One-loop vaccum amplitudes of the  6D model in the transverse channel}
The one-loop  Klein bottle amplitude  in the transverse channel is given by :
\bea
\tilde{\cal K}&=&\tilde{\cal K}_0+\frac{2^5}{4}(\sqrt{v_4})^2\left[\sum_n 
\tilde{Z}_{2n}\tilde{\Lambda}_e^{(3)}\right]'(V_4I_4+I_4V_4-S_4S_4-C_4C_4)
\nonumber \\&&
+\frac{2^5}{4}(\frac{1}{\sqrt{v_4}})^2\left[\sum_n Z_{2n}\Lambda_e^{(3)}\right]'
(V_4I_4+I_4V_4)\nonumber \\&&-\frac{2^5}{4}(\frac{1}{\sqrt{v_4}})^2
\left[\sum_nZ_{2n+1}\Lambda_e^{(3)}\right]'(S_4S_4+C_4C_4)\nonumber
\eea
with $\left[\,\,\,\right]'$ means that the Narain's lattice is not included. The $v_4$ is the volume of the compact space, in $\Lambda_e^{(3)}$ and
${\tilde \Lambda}_e^{(3)}$ the lattice sums are restricted to even values of 
momenta and winding, respectively. The contribution $\tilde{\cal K}_0$ from the origin of Narain lattice is : 
\bea
\tilde{\cal K}_0&=&\frac{2^5}{4}(\sqrt{v_4}+\frac{1}{\sqrt{v_4}})^2(V_4I_4(I_4I_4)_B+I_4V_4(V_4V_4)_B)\nonumber \\&&+\frac{2^5}{4}(\sqrt{v_4}-\frac{1}{\sqrt{v_4}})^2(V_4I_4(V_4V_4)_B+I_4V_4(I_4I_4)_B)\nonumber \\&&-\frac{2^5}{4}(\sqrt{v_4})^2(S_4S_4+C_4C_4)(I_4I_4+V_4V_4)_B\nonumber
\eea
where the characters with label $B$  correspond to:
\bea
(I_4I_4+V_4V_4)_B&=&\left[\sum_n \tilde{Z}_{2n}\tilde{\Lambda}_e^{(3)}\right]_0=\, \left[\sum_n Z_{2n}\Lambda_e^{(3)}\right]_0, \nonumber \\ 
(I_4I_4-V_4V_4)_B&=&4\frac{\eta^2}{\theta_2^2},\qquad(Q_S+Q_C)_B=4\frac{\eta^2}{\theta_4^2},\qquad (Q_S-Q_C)_B=4\frac{\eta^2}{\theta_3^2}\nonumber
\eea
In the transverse channel, the annulus amplitude is :
\bea
2^7\tilde{\cal A}&=&2^7\tilde{\cal A}_0+(\sqrt{v_4})^2n_N^2(V_4I_4+I_4V_4-S_4S_4-C_4C_4)\left[\sum_n\tilde{Z}_{2n}\tilde{\Lambda}^{(3)}\right]'\nonumber \\&&+(\sqrt{v_4})^2n_N^2(I_4I_4+V_4V_4-S_4C_4-C_4S_4)\left[\sum_n\tilde{Z}_{2n+1}\tilde{\Lambda}^{(3)}\right]'\nonumber \\&&+(\frac{1}{\sqrt{v_4}})^2(n_{D_1}+n_{D_2})^2(V_4I_4+I_4V_4)\left[\sum_n\tilde{Z}_{2n}\tilde{\Lambda}^{(3)}\right]'\nonumber \\&&-(\frac{1}{\sqrt{v_4}})^2(n_{D_1}-n_{D_2})^2(S_4S_4+C_4C_4)\left[\sum_n\tilde{Z}_{2n}\tilde{\Lambda}^{(3)}\right]'\nonumber \\&&+(\frac{1}{\sqrt{v_4}})^2(n_{D_1}-n_{D_2})^2(V_4I_4+I_4V_4)\left[\sum_n\tilde{Z}_{2n+1}\tilde{\Lambda}^{(3)}\right]'\nonumber \\&&-(\frac{1}{\sqrt{v_4}})^2(n_{D_1}+n_{D_2})^2(S_4S_4+C_4C_4)\left[\sum_n\tilde{Z}_{2n+1}\tilde{\Lambda}^{(3)}\right]'\nonumber
\eea
where  $\tilde{\cal A}_0$ corresponds to the contribution from 
the origin of the Narain's lattice :
\bea
2^7\tilde{\cal A}_0&=&(\sqrt{v_4}n_N+\frac{n_{D_1}+n_{D_2}}{\sqrt{v_4}})^2(V_4I_4(I_4I_4)_B+I_4V_4(V_4V_4)_B)\nonumber \\&&+(\sqrt{v_4}n_N-\frac{n_{D_1}+n_{D_2}}{\sqrt{v_4}})^2(V_4I_4(V_4V_4)_B)+I_4V_4(I_4I_4)_B)\nonumber \\&&-(\sqrt{v_4}n_N+\frac{n_{D_1}-n_{D_2}}{\sqrt{v_4}})^2(C_4C_4(I_4I_4)_B+S_4S_4(V_4V_4)_B)\nonumber 
\eea
\bea
\qquad &&-(\sqrt{v_4}n_N-\frac{n_{D_1}-n_{D_2}}{\sqrt{v_4}})^2(C_4C_4(V_4V_4)_B+S_4S_4(I_4I_4)_B)\nonumber \\&&+\left[\frac{(R_N+4R_{D_1})^2}{4}+\frac{7R_N^2}{4}\right](Q_SQ_{SB}+Q_CQ_{CB})\nonumber \\&&+\left[\frac{(R_N-4R_{D_1})^2}{4}+\frac{7R_N^2}{4}\right](Q_SQ_{CB}+Q_CQ_{SB})\nonumber \\&&+\left[\frac{(R_N+4R_{D_2})^2}{4}+\frac{7R_N^2}{4}\right](Q'_SQ_{SB}+Q'_CQ_{CB})\nonumber \\&&+\left[\frac{(R_N-4R_{D_2})^2}{4}+\frac{7R_N^2}{4}\right](Q'_SQ_{CB}+Q'_CQ_{SB})
\nonumber
\eea
Furthermore, for the M{\"o}bius transverse channel, we found :
\bea
-2\tilde{\cal M}&=&-2\tilde{\cal M}_0+n_Nv_4(\hat{V}_4\hat{I}_4+\hat{I}_4\hat{V}_4)\left[\sum_n(\tilde{Z}_{4n}+\tilde{Z}_{4n+2})\tilde{\Lambda}_e^{(3)}\right]'\nonumber \\&&-n_Nv_4(\hat{S}_4\hat{S}_4+\hat{C}_4\hat{C}_4)\left[\sum_n(\tilde{Z}_{4n}-\tilde{Z}_{4n+2})\tilde{\Lambda}_e^{(3)}\right]'\nonumber \\&&+\frac{n_{D_1}+n_{D_2}}{v_4}(\hat{V}_4\hat{I}_4+\hat{I}_4\hat{V}_4)\left[\sum_nZ_{2n}\Lambda_e^{(3)}\right]'\nonumber \\&&+\frac{n_{D_1}+n_{D_2}}{v_4}(\hat{S}_4\hat{S}_4+\hat{C}_4\hat{C}_4)\left[\sum_n Z_{2n+1}\Lambda_e^{(3)}\right]'\nonumber
\eea
with the contribution from the origin $\tilde{\M}_0$  given by :
\bea
-2\tilde{\M}_0&=&(\sqrt{v_4}+\frac{1}{\sqrt{v_4}})
(\sqrt{v_4}n_N+\frac{n_{D_1}+n_{D_2}}{\sqrt{v_4}})
(\hat{V}_4\hat{I}_4(\hat{I}_4\hat{I}_4)_B+
\hat{I}_4\hat{V}_4(\hat{V}_4\hat{V}_4)_B)\nonumber \\
&&+(\sqrt{v_4}-\frac{1}{\sqrt{v_4}})(\sqrt{v_4}n_N-\frac{n_{D_1}
+n_{D_2}}{\sqrt{v_4}})(\hat{V}_4\hat{I}_4(\hat{V}_4\hat{V}_4)_B)
+\hat{I}_4\hat{V}_4(\hat{I}_4\hat{I}_4)_B))\nonumber \\
&&-\sqrt{v_4}(\sqrt{v_4}n_N-\frac{n_{D_1}-n_{D_2}}{\sqrt{v_4}})
(\hat{C}_4\hat{C}_4(\hat{V}_4\hat{V}_4)_B+\hat{S}_4\hat{S}_4
(\hat{I}_4\hat{I}_4)_B)\nonumber \\
&&-\sqrt{v_4}(\sqrt{v_4}n_N+\frac{n_{D_1}-n_{D_2}}{\sqrt{v_4}})
(\hat{C}_4\hat{C}_4(\hat{I}_4\hat{I}_4)_B+\hat{S}_4\hat{S}_4
(\hat{V}_4\hat{V}_4)_B)\nonumber \, .
\eea

\end{document}